\documentclass{article}

\usepackage{amssymb,amsmath}     
\usepackage{graphicx, tikz}    
\usepackage{amsthm}      
\usepackage{combgames}
\usepackage{enumerate}
 \usepackage{natbib}
 \usepackage{mathrsfs}

 \usetikzlibrary{arrows.meta}

\usepackage[hypertexnames=false]{hyperref}
\newcommand{\cg}[2]{\left\{#1\,\middle|\,#2\right\}}

\theoremstyle{plain}
\newtheorem{thm}{Theorem}[section]
\newtheorem{cor}[thm]{Corollary}
\newtheorem{lem}[thm]{Lemma}
\newtheorem{prop}[thm]{Proposition}
\newtheorem{conj}[thm]{Conjecture}

\theoremstyle{definition}
\newtheorem{defi}[thm]{Definition}

\newtheorem{ex}[thm]{Example}
\newtheorem{problem}[thm]{Problem}
\newtheorem{obs}[thm]{Observation}

\renewcommand{\L}{\mathrm L}
\newcommand{\R}{\mathrm R}
\newcommand{\N}{\mathrm N}
\renewcommand{\P}{\mathrm P}

\newcommand{\zug}{\ensuremath{\mathrm Z}}

\newcommand{\nL}{\ensuremath{\mathscr L}}
\newcommand{\nR}{\ensuremath{\mathscr R}}
\newcommand{\nN}{\ensuremath{\mathscr N}}
\newcommand{\nP}{\ensuremath{\mathscr P}}

\newcommand{\GL}{G^{\mathcal L}}
\newcommand{\GR}{G^{\mathcal R}}
\newcommand{\HL}{H^{\mathcal L}}
\newcommand{\HR}{H^{\mathcal R}}

\title{A number game reconciliation}
\author{Prem Kant and Urban Larsson}
\date{June 2025}

\begin{document}

\maketitle
\begin{abstract}
Number games play a central role in alternating normal play combinatorial game theory due to their real-number-like properties (Conway 1976). 
Here we undertake a critical re-examination:  we begin with integer and dyadic games and identify subtle inconsistencies and oversights in the established literature (e.g. Siegel 2013), most notably, the lack of distinction between a game being a number and a game being equal to a number. After addressing this, we move to the general theory of number games. We analyze Conway’s original definition and a later refinement by Siegel, and highlight conceptual gaps that have largely gone unnoticed. Through a careful dissection of these issues, we propose a more coherent and robust formulation. Specifically, we develop a refined characterization of numbers, via several subclasses,  dyadics, canonical forms, their group theoretic closure and zugzwangs, that altogether better capture the essence of number games. This reconciliation not only clarifies existing ambiguities but also uncovers several open problems.
\end{abstract}

\section{Introduction}

Combinatorial Game Theory (CGT) involves two-player games of perfect information and no chance. In alternating normal play, the players take turns making moves, and the game ends when a player cannot move on their turn; that player loses. In other words, the player who makes the last move wins.

\begin{figure}[h!]
\centering
\includegraphics[width=0.27\textwidth]{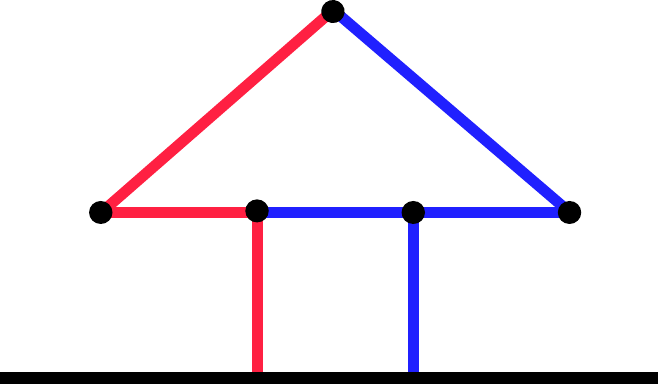}
\caption{A {\sc Hackenbush} position.}
\label{fig:hackenbush}
\end{figure}

A classic example is Hackenbush. A simple game position is shown in Figure~\ref{fig:hackenbush}. In this game, the two players, Red and Blue, take turns removing red and blue edges, respectively. An edge may be removed only if it is connected to the ground. When an edge is removed, any edges that are no longer connected to the ground automatically fall. The game naturally ends when no more removable edges remain.


In the position shown in Figure~\ref{fig:hackenbush}, the moves available to each player depend only on their own previous choices, not on the opponent's. Blue has four available moves, while Red has three. Regardless of who starts, Blue wins under optimal play. Notably, this position can be interpreted as one where Blue effectively has one move and Red has none. In CGT, we refer to such a position as the number game ``$1$''. More generally, every Hackenbush position can be assigned a numerical value that captures its strategic essence.


\begin{figure}[ht!]
        \centering
        \includegraphics[width=0.2\linewidth]{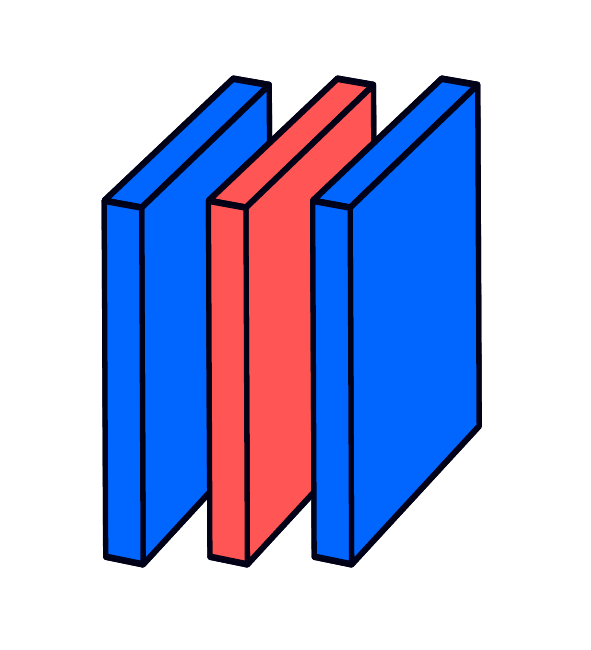}
        \caption{\small A game of \textsc{Toppling Dominoes} with two blue pieces and a red piece in the middle.}
        \label{fig:halftoppl}
\end{figure}

But numbers are not unique to Hackenbush; they appear in many other combinatorial games as well. For example, consider the game of \textsc{Toppling Dominoes}, illustrated in Figure~\ref{fig:halftoppl}. In this game, the two players, Left (b{\bf L}ue) and Right ({\bf R}ed), take turns toppling dominoes of their respective colors in either direction. The position shown in Figure~\ref{fig:halftoppl} corresponds to the dyadic number $1/2$, providing another example of how numerical values naturally arise in combinatorial game theory.


Number games play a central role in CGT due to their real number-like properties~\cite{ONAG}. The conceptual foundations for these games were laid by John Conway in his seminal treatise On Numbers and Games~\cite{ONAG}, which shaped much of the modern development of the subject. In Conway's framework, a number is defined as a game in which no Left option is greater than or equal to any Right option, and every option is itself a number. This elegant condition allows for the construction of all real numbers, and much more.

However, this definition contains a subtle inconsistency that has been largely overlooked: there exist game positions that satisfy the formal definition of a number, yet their equivalent positions do not. In other words, ``being a number'' is not preserved under equivalence. This inconsistency is troubling, since CGT treats equivalence classes of games---not individual representatives---as fundamental. It suggests that the current definition of number is not fully compatible with the foundational principles of the theory.

Aaron Siegel, in his book Combinatorial Game Theory~\cite{S2013}, proposed a modified definition. He defines a number as a game in which every Left option is strictly less than every Right option, and all options are themselves numbers. While this revision strengthens the original condition, the underlying ambiguity remains: there still exist games that satisfy this definition but are equivalent to positions that do not. Thus, the issue that ``being a number is not preserved under equivalence'' persists even under Siegel’s refinement.


We believe the source of this issue lies in defining numbers for equivalence classes of games by considering only their canonical forms, rather than their literal forms. As a result, certain literal-form games do not satisfy the formal definition of numbers, even though their canonical forms do. This subtle but important distinction has so far been overlooked in the literature.

To address this issue, in this paper, we revisit the theory of numbers and their constituents, integers, dyadics, and zugzwangs, restricting our attention to games that are \textit{short}. Unlike the approach of~\citep[Definition 3.4, Theorem 3.6]{S2013}, which focuses on equivalence classes and canonical forms, we work directly with the set of literal game forms. This allows us to highlight the existing inconsistencies and provide a corrected, consistent framework for defining numbers.






In Section~\ref{sec:basicsetup}, we review the basic setup and present preliminary results. In Section~\ref{sec:integeralt}, we revisit the theory of integer games. Section~\ref{sec:dyadicalt} introduces dyadic games, highlights gaps in the existing literature, and resolves them. In the process, we present a more explicit version of the canonical forms of dyadic games and also prove their dense-in-itself property. Section~\ref{sec:numbersalt} discusses Conway numbers and Siegel numbers, points out inconsistencies, and establishes their equivalence with zugzwang positions. We define numbers as the full equivalence class of dyadic games and provide an alternate characterization based on the notion of a ``fitting set''. We conclude this section with the well-known Archimedean property and the number avoidance theorem. Finally, in Section~\ref{sec:infinitesimalsalt}, we present results on weak zugzwangs, infinitesimals, and outline open problems.

\section{Basic Set up and Preliminary Results}\label{sec:basicsetup}

In this section, we set up the basic framework and review the key results from combinatorial game theory that will be used throughout the paper.

\subsection{Basic set up}\label{subsec:basicsetup}

Throughout this paper, we work with \textit{short} combinatorial games. A game is called \textit{short} if no position can be repeated, and from every position, each player has only finitely many possible moves. As is standard, the two players are called Left (female) and Right (male).

The \textbf{game form} of a short game $G$ is defined recursively as $G = \cg{\GL}{\GR}$, where $\GL = \{G_1^L, G_2^L, \ldots, G_m^L\}$ is the finite set of all Left options of $G$, and $\GR = \{G_1^R, G_2^R, \ldots, G_n^R\}$ is the finite set of all Right options of $G$. Instead of writing $G = \cg{\GL}{\GR}$, we usually list all options explicitly and write $G=\cg{G^L_1, G^L_2, \ldots, G^L_m}{G^R_1, G^R_2, \ldots, G^R_n}$. The set of all short games is denoted by $\mathbb{G}$.

A game $G$ is called \textit{terminal} for a player if that player has no available moves. Specifically, if $\GL = \varnothing$, Left cannot move; if $\GR = \varnothing$, Right cannot move. If both sets are empty, that is, $\GL = \GR = \varnothing$, then $G$ is terminal for both players and is referred to as the \textbf{zero game}, denoted $G = 0$.

We refer to a typical Left (Right) option of $G$ as $G^L$ ($G^R$). A game form $H$ is a \textit{subposition} or \textit{follower} of $G$ if there exists a finite sequence of moves, by either player, not necessarily alternating, that takes $G$ to $H$.

\begin{figure}[ht!]
        \centering
        \includegraphics[width=\linewidth]{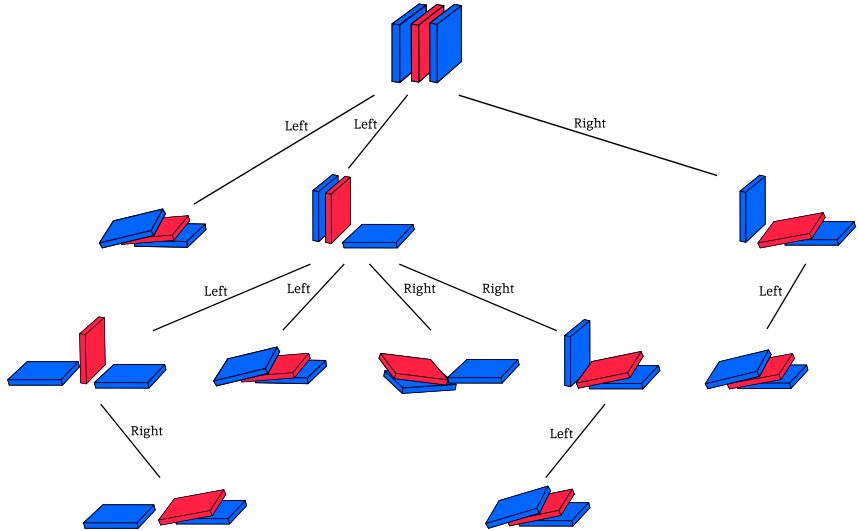}
       \caption{\small The game tree of the \textsc{Toppling Dominoes} position of Figure~\ref{fig:halftoppl}, up to symmetry.}
        \label{fig:gametreehalftoppl}
\end{figure}

With the game form already defined, we can now visualize the combinatorial structure of a game $G$ using a \textbf{game tree}.  This is a directed graph in which each node represents a subposition $H$ of $G$, with  \textit{left slanting edges} to each Left option $H^L$ and  \textit{right slanting edges } to each Right option $H^R$. For example, see Figure~\ref{fig:gametreehalftoppl}, which shows the game tree of the \textsc{Toppling Dominoes} position in Figure~\ref{fig:halftoppl}. Using backward induction on this game tree, the game form of the position is obtained as
$$\cg{0,\cg{\cg{\varnothing}{0},0}{0,\cg{0}{\varnothing}}}{\cg{0}{\varnothing}}.$$  
To simplify notation, we can omit explicitly writing ``$\varnothing$'' in the game form and express it as 
$$\cg{0,\cg{\cg{~}{0},0}{0,\cg{0}{~}}}{\cg{0}{~}}.$$
This fully expanded representation, where all subpositions are explicitly mentioned, is called the \textbf{literal form} of the game.

\subsection{Algebra of Games}
We now discuss the algebraic structure of combinatorial games.

\begin{defi}[Disjunctive Sum] \label{def:disjunctivesum} Let $G=\cg{\GL}{\GR}$ and $H=\cg{\HL}{\HR}$ be two games in $\mathbb G$. The disjuctive sum $G+H$ is defined recursively as: $$G+H=\cg{G+\HL,\GL+H}{G+\HR, \GR+H},$$ where $G+\HL=\{G+H^L: H^L\in\HL\}$, in case $\HL\neq\varnothing$, and otherwise the set is not defined and omitted (commas between options indicate set union). Other terms are defined similarly.
\end{defi}

As defined, the disjunctive sum satisfies both commutativity and associativity. We now define the \textit{conjugate} of a game form. 
\begin{defi}[Conjugate] For any game $G=\cg{\GL}{\GR}$, the  conjugate $\overline G$ is defined recursively as: $$\overline G=\cg{\overline \GR}{\overline \GL}.$$
\end{defi}

We next define the equality relation  ``$=$'' between games. Here  we denote the \textbf{outcome class} of a game $G$ by $o(G)$.
\begin{defi}
Consider $G, H \in \mathbb G$. Then $G=H$ if and only if, for every game $X \in \mathbb G$, $o(G+X) = o(H+X)$. 
\end{defi}

The relation ``$=$'' is fundamental to the theory of combinatorial games; therefore, it is essential that it is well defined. In this context, the fundamental theorem by Zermelo guarantees that, for a  given starting player, every combinatorial game has a unique outcome under perfect play. This ensures that the equality ``$=$'' is indeed well defined, as no position can lead to multiple distinct results.

It is easy to verify that ``$=$'' is an equivalence relation. However, $G = H$ does not necessarily imply that $G$ and $H$ have the same game form. When two games have the same game form, we say they are \emph{identical} and write $G \cong H$.

Partizan games have four distinct outcome classes: Left wins regardless of who starts ($\nL$), Right wins regardless of who starts ($\nR$), the starting player wins ($\nN$), and the second player wins ($\nP$). With a slight abuse of notation, we write:
$\nL = \L\L, \,\nN = \L\R, \,\nP = \R\L$, and $ \nR = \R\R$.
Here, ``$\L\R$'' means that Left wins when Left plays first, and Right wins when Right plays first, while ``$\R\L$'' means that Right wins when Left starts, and Left wins when Right starts. The meanings of ``$\L\L$'' and ``$\R\R$'' are analogous.

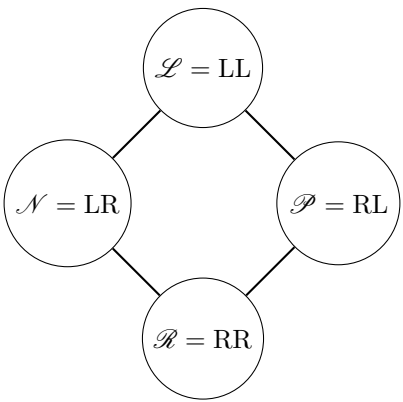
\begin{figure}
\centering{
\begin{tikzpicture}[scale = 1.8]
\begin{scope}[every node/.style={circle, draw}]  
    \node (2) at (0,0) {$\nP= \R\L$};
    \node (4) at (-1,-1) {$\nR= \R\R$};
    \node (0) at (-1,1) {$\nL = \L\L$};
    \node (1) at (-2,0) {$\nN = \L\R$};
    \draw (0, 1) {};
\end{scope}
\begin{scope}[>={Stealth[black]},
              every node/.style={fill=white,circle},
              every edge/.style={draw=black,thick}] 
    \path [-] (1)edge(0);
    \path [-] (2)edge(0);
    \path [-] (1) edge (4);
    \path [-] (2) edge (4);
\end{scope}
\end{tikzpicture}}
\caption{The outcome diamond}
\label{diamond}
\end{figure}

Adopting the convention of Left favorability, we define the \textit{outcome relation} as follows:

\begin{defi}[Outcome Relation] The outcomes $\nL=\L\L, \,\nN = \L\R, \,\nP = \R\L$, and $ \nR  =\R\R$ satisfy the following:
\begin{enumerate}[i)]
    \item $\L\L>\L\R>\R\R$,
    \item $\L\L>\R\L>\R\R$, 
    \item $\L\R \cgfuzzy\R\L$.
\end{enumerate}
\end{defi}
By definition, these outcomes are naturally partially ordered with relation ``$\ge$''. Moreover, they form a lattice. The Hasse diagram of these outcomes, known as the \textit{outcome diamond}, is shown in Figure~$\ref{diamond}$. 

With the outcome relation defined, we now introduce the \textit{game relation}.

\begin{defi}[Game Relation]\label{def:gamerelalt}
Consider games $G, H \in \mathbb G $. Then  $G \ge H$ if, for all  $X \in \mathbb G$,  $o(G + X) \ge o(H + X)$.
\end{defi}

The game relation ``$\ge$'' induces a partial order on  $\mathbb{G}$.  It is also translation-invariant, making $\mathbb G$ a partially ordered group.

We have already seen the zero game, $0 = \cg{}{}$. In combinatorial game theory, the zero game is the identity element, whose equivalence class consists of all games $G$ with $o(G) = \nP$. It is also considered the neutral game and acts as the reference point for comparing games, as summarized in Table~\ref{tbl:rlnG0}.

\begin{table}[ht]
\centering
\renewcommand{\arraystretch}{1.5}
\caption{Relation between games $G$ and $0$.}
\begin{tabular}{|l|}
\hline
$G = 0\iff o(G) = \nP \iff$  second player always wins\\ \hline
$G > 0 \iff o(G)  = \nL \iff$ Left always wins\\ \hline
$G < 0 \iff o(G) = \nR \,\iff$  Right always wins\\ \hline
$G \cgfuzzy 0 \,\iff o(G) = \nN \!\iff$ first player always wins\\ \hline
$G \ge 0 \iff o(G) \ge \nP \iff$ Left wins playing second \\ \hline
$G \le 0 \iff o(G) \le \nP \iff$ Right wins playing second \\ \hline
$G \cggfuz 0 \!\iff o(G) \ge \nN \!\iff$ Left wins playing first \\ \hline
$G \cglfuz 0 \!\iff o(G) \le \nN \!\iff$ Right wins playing first \\
\hline
\end{tabular}
\label{tbl:rlnG0}
\end{table}

We now state an important result that describes the relation between any game $G$ and its options.

\begin{lem}\label{lem:GLGGR}
    Consider a game $G = \cg{G^{\mathcal L}}{G^{\mathcal R}} \in \mathbb G$. For every $G^L \in G^{\mathcal{L}}$ and $G^R \in G^{\mathcal R}$, we have $G^L \cglfuz G \cglfuz G^R$.
\end{lem}

\begin{proof}
    For any $G^L \in G^{\mathcal L}$, consider the game $G^L - G$. If Right starts, he plays to $G^L - G^L = 0$ and wins. Therefore, $G^L - G \cglfuz 0$, which implies $G^L \cglfuz G$. 

    Similarly, for any $G^R \in G^{\mathcal R}$, we get $G\cglfuz G^R$.
\end{proof}

\subsection{Canonical form}
Recall the \textsc{Toppling Dominoes} game from Figure~\ref{fig:halftoppl}. In Subsection~\ref{subsec:basicsetup}, we saw that its literal game form is
$$ \cg{0,\cg{\cg{~}{0},0}{0,\cg{0}{~}}}{\cg{0}{~}}. $$ 
 This is equivalent to the simplified form 
$$ 1/2 = \cg{0}{\cg{0}{~}},$$
where the inferior Left option has been eliminated.

This simplification is possible by the notions of \emph{domination} and \emph{reversibility}, which allow the removal of suboptimal or reversible options from a game form.
We now define these notions precisely:

\begin{defi}[Dominated and Reversible options] 
Consider a game $G \in \mathbb G$.
\begin{enumerate} [i)]
    \item A Left option $G^{L_1}$ is dominated by another Left option $G^{L_2}$ if $G^{L_2} \ge G^{L_1}$.
    \item A Right option $G^{R_1}$ is dominated by another Right option $G^{R_2}$ if $G^{R_2} \le G^{R_1}$.
    \item A Left option $G^{L_1}$ is reversible through $G^{L_1R_1}$ if $G^{L_1R_1} \le G$.
    \item A Right option $G^{R_1}$ is reversible through $G^{R_1L_1}$ if $G^{R_1L_1} \ge G$.
\end{enumerate}
\end{defi}

The removal of dominated and reversible options does not change the equivalence class of the game.

\begin{thm}\label{thm:reductionalt}
Consider a game $G = \cg{G^{\mathcal L}}{G^{\mathcal R}} \in \mathbb G$. 
\begin{enumerate}[i)]
   \item \textbf{Domination:}  If there exist Left options $G^{L_1}, G^{L_2} \in G^{\mathcal L}$ such that $G^{L_2} \ge G^{L_1}$, then 
$$G = \cg{G^{\mathcal L}\setminus\{{G^{L_1}}\}}{G^{\mathcal R}}.$$
\item \textbf{Reversibility:} If there exists a Left option $G^{L_1} \in G^{\mathcal L} $ with a Right response $G^{L_1R_1}$ such that $G^{L_1R_1} \le G$, then $$G = \cg{G^{\mathcal L} \setminus\{G^{L_1}\}, G^{L_1R_1\mathcal L}}{G^{\mathcal R}}.$$
\end{enumerate}
\end{thm}

Similarly, any dominated or reversible Right option can be eliminated when simplifying the game form.

Given a game $G \in \mathbb{G}$, we can apply Theorem~\ref{thm:reductionalt} repeatedly to reduce $G$ to a simpler form. The final result, after no further reductions are possible, is called the \emph{canonical form} of $G$.

\begin{defi}[Canonical Form]
Consider a game $G \in \mathbb{G}$. A game $H$ is said to be the \emph{canonical form} of $G$ if:
\begin{enumerate}
    \item $G = H$, and
    \item No subposition of $H$ contains any dominated or reversible options.
\end{enumerate}
\end{defi}

Regardless of the order in which Theorem~\ref{thm:reductionalt} is applied to simplify $G$, the resulting canonical form is unique. The canonical form of a game $G$ is often also called its \emph{game value}.


\section{Integer games}\label{sec:integeralt}

We now begin our main investigation with the class of games corresponding to integers. These games represent the \emph{move advantage} held by one of the players. For example, the integer game ``$3$'' means that Left has three moves while Right has no moves available, whereas the game ``$-2$'' is worth exactly two moves for Right. Although the results and proofs in this section are well known, we include them for completeness.

To distinguish between the integer $n$ and the corresponding game form in this section, we use the notation $(n)$ to denote the integer game form.

\begin{defi}[Integer Games]\label{def:integergames} 
The game corresponding to the integer $0$ is defined as 
$(0) = \cg{\varnothing}{\varnothing}$.
For each $n \in \mathbb{N} = \{1, 2, 3, \ldots\}$, the game corresponding to the integer $n$ is recursively defined as
$(n) = \cg{(n-1)}{\varnothing}$.
The game $(-n)$ is defined as the inverse of $(n)$. 
These literal form games constitute the set of integer games, denoted by $\mathbb{I}$, given by
$ \mathbb{I} = \{(n) : n \in \mathbb{Z}\} $.
\end{defi}

In a gameplay scenario, the game $(n)$ can be interpreted as $n$ free moves for Left, and similarly, $(-n)$ corresponds to $n$ free moves for Right.

We now prove some properties of integer games.
\begin{prop}\label{prop:Lwinnalt}
For any $n \in \mathbb{N}$, the game form $(n) = \cg{(n-1)}{\varnothing}$ satisfies $(n) > (0)$. 
\end{prop}
\begin{proof}
    For $n \in \mathbb{N}$, consider the game form $(n) = \cg{(n-1)}{\varnothing}$.  If Left starts, she plays to $(n-1)$. Now it is Right's turn, but he has no available moves and therefore loses.  If Right starts, he has no move and immediately loses. Thus, Left always wins the game $(n)$, and hence $(n) > (0)$.
\end{proof}
Proposition~\ref{prop:Lwinnalt} leads to the following observation about the inverses of positive integer games.
\begin{obs}\label{obs:Rwinnalt}
  For any $n \in \mathbb{N}$, by Proposition~\ref{prop:Lwinnalt} we have $(n)>(0)$. Therefore, its inverse $(-n)$ satisfies  $(-n) = \cg{\varnothing}{(-n+1)}<(0)$.
\end{obs}

We now prove that the disjunctive sum of two integer game forms is again equal to an integer game.

\begin{thm}\label{thm:integersaddalt} Consider $(n), (m) \in \mathbb I$. Then the disjunctive sum 
$(n) + (m) = (n + m)$.
\end{thm}
\begin{proof}
If either of the integers $n$ or $m$ is $0$, then the result follows immediately.

\noindent \textbf{Case 1:} Suppose $n, m > 0$. Then $(n) = \cg{(n-1)}{\varnothing}$ and $(m) = \cg{(m-1)}{\varnothing}$. In the sum $(n) + (m)$, Right has no available moves. Left can move to either $(n-1)+(m)$ or $(n)+(m-1)$. Let $(k) = \cg{(n+m-1)}{\varnothing}$. By induction, we have $(n-1)+(m) = (n+m-1)$ and $(n)+(m-1) = (n+m-1)$. Hence, the options of $(n)+(m)$ match those of $(k)$, and therefore $(n)+(m) = \cg{(n+m-1)}{\varnothing} = (n+m)$.

\noindent \textbf{Case 2:} The case when both $n, m < 0$ is symmetric.

\noindent \textbf{Case 3:} Assume $n>0$ and $m<0$.

 \noindent \textbf{Subcase a:} \textbf{$n< -m$} 
\begin{align*}
    (n)+(m) &= (n) + (m+n-n) &&\\
    &= (n) + (- n) + (m+ n)&& \text{since } (-n)<0 \text{ and } (m+n)<0\\
    &= (m+ n)&& \text{since }(-n) \text{ is the inverse of } (n)
\end{align*}
\noindent \textbf{Subcase b:} \textbf{$n>- m$} 
\begin{align*}
    (n)+(m) &= (n+m-m) + (m) &&\\
    &=(n+m) + (-m) + (m)&& \text{since } -m>0 \text{ and } (n+ m)>0\\
    &= (n+ m)&&
\end{align*}
\noindent \textbf{Subcase c:} \textbf{$n=- m$}
$$(n)+(m)=(0) = (n+m).$$
This completes the proof.
\end{proof}

Note that the game form of the disjunctive sum $(n) + (m)$ is not necessarily identical to the game form of $(n + m)$. For example, consider the games $(1) = \cg{(0)}{\varnothing}$ and $(-1) = \cg{\varnothing}{(0)}$. Their disjunctive sum is
\begin{equation}\label{eq:integernotinI}
    (1) + (-1) \cong \cg{-1}{1},
\end{equation}
whereas $(1 - 1) = (0) = \cg{\varnothing}{\varnothing}$. Therefore, $(1) + (-1) \ncong (0) = (1 - 1)$. Thus, the set $\mathbb{I}$ is not closed under disjunctive sum. 

We therefore define the set $\widetilde {\mathbb I}$, consisting of all game forms that are equal to integer games:
$$\widetilde{\mathbb I} =\{ G \in \mathbb G \mid G=(x) \text{ for some } (x) \in \mathbb I \}.$$
We now show that $\widetilde{\mathbb I}$ is a totally ordered group.
\begin{prop}\label{prop:tildeintegergroup}
    The set $\widetilde{\mathbb I}$ is a totally ordered group under the game relation ``$\ge$''.
\end{prop}
\begin{proof}
    We need to prove the following:
    \begin{enumerate}[i)]
        \item $\widetilde{\mathbb I}$ is a group.
        \item The game relation ``$\ge$'' is a total order on $\widetilde{\mathbb I}$.
        \item The game relation ``$\ge$'' is translation-invariant on $\widetilde{\mathbb I}$.
    \end{enumerate}
Let $G, H \in \widetilde{\mathbb I}$. By definition, there exist integer games $(n)$ and $(m) $ such that $G=(n)$ and $H = (m)$.\\

 \noindent   \textit{Proof of i).} Since, $\widetilde{\mathbb I} \subset \mathbb G$ and $\mathbb G$ is a  group, it suffices to show that $G-H \in \widetilde{\mathbb I}$. 
    
    By Theorem~\ref{thm:integersaddalt}, $G-H = (n) - (m) = (n-m)$. Since $(n-m) \in \mathbb I$,  it follows that $G-H \in \widetilde{\mathbb I}$. Thus, $\widetilde{\mathbb I}$ is a subgroup of $\mathbb G$.\\

  \noindent  \textit{Proof of ii).} Since $\mathbb Z$ is totally ordered, we have either $n \ge m$ or $m \ge n$. This implies either $n-m \ge 0$ or $m-n \ge 0$. If $n-m \ge 0$, then by Proposition~\ref{prop:Lwinnalt}, we have $(n-m) \ge (0)$. By Theorem~\ref{thm:integersaddalt} this implies $(n) + (-m) \ge 0$. Thus, $(n) \ge (m)$ and hence $G\ge H$. Similarly,   if $m - n \ge 0$, we have $(m) \ge (n)$, and hence $H \ge G$.

    This shows that the relation ``$\ge$'' is a total order on $\widetilde{\mathbb I} $.\\ 

   \noindent \textit{Proof of iii).} Since the game relation ``$\ge$'' is translation-invariant on $\mathbb{G}$, it remains translation-invariant on the subclass $\widetilde{\mathbb{I}}$ as well.
 \end{proof}
Since $\mathbb I \subset \widetilde{\mathbb I}$, Proposition~\ref{prop:tildeintegergroup} implies that  $\mathbb I$ is totally ordered under the same relation ``$\ge$''.

We conclude this section by presenting an alternative expression for integer games.

\begin{lem}\label{lem:intsimplalt}
For all integers $n \in \mathbb Z$, $(n)=\cg{(n-1)}{(n+1)}$.
\end{lem}

\begin{proof}
Let us define the game $G := \cg{(n-1)}{(n+1)} +(- n)$. It is sufficient to show that $G = (0)$.

\noindent \textbf{Case 1:} $-n \ge 1$. Then,
\begin{align*}
    G &= \cg{(n-1)}{(n+1)} + (- n)\\
    &= \cg{(n-1)}{(n+1)} + \cg{(-n-1)}{\varnothing}.
\end{align*}

 If Left starts, she plays to either $(n-1)+(-n)$ or to $\cg{(n-1)}{(n+1)} + ( - n-1)$. She loses  $(n-1)+(-n) = (n-1-n) = (-1)$ as it is a negative game. In the other game $\cg{(n-1)}{(n+1)} +(-n-1)$, Right plays to $(n+1)+(-n-1)=(n+1-n-1)= (0)$ and wins.

 If Right starts, he can only play to $(n+1)+(-n) = (1)$, which he loses.

\noindent \textbf{Case 2:} $n=0$. Then $G= \cg{(-1)}{(1)}$.

If Left starts, she can only play to $(-1)$, which she loses.

If Right starts, he can only play to $(1)$, which he loses.

\noindent \textbf{Case 3:} $-n \le -1$. Then $G = \cg{(n-1)}{(n+1)} + \cg{\varnothing}{(-n+1)}$. 

If Left starts, she plays to $(n-1)+(-n) = (-1)$ and loses.

If Right starts, he plays to either $(n+1)+(-n)$ or to $\cg{(n-1)}{(n+1)} +(-n+1)$. Left wins $(n+1)+(-n) = (1)$ as it is a positive game. In the game $\cg{(n-1)}{(n+1)}+(-n+1)$, Left plays to $(n-1)+(-n+1) = (0)$ and wins.

Hence, the second player wins the game $G$. Therefore, $G=(0)$.
\end{proof}

By a similar proof to that of Lemma~\ref{lem:intsimplalt}, we can get that for any $n, k \in \mathbb{N}$, the game $\cg{(n-1)}{(n+k)}$ is equal to the integer $(n)$.

\section{Dyadic rational games}\label{sec:dyadicalt}
We begin by defining a class of games that correspond to dyadic rational numbers.

\begin{defi}[Dyadic Rational Games]\label{def:dyadicgames}
The game $1/2^0 = 1$.
For each $k \in \mathbb{N}$, 
let $1/2^k=\cg{0}{1/2^{k-1}}$.
More generally, for any $m \in \mathbb{N}$, the game $m/2^k$ 
is the disjunctive sums of $m$ copies of $1/2^k$, and  $-m/2^k$ is the  disjunctive sums of $m$ copies of $-1/2^k$. 
These literal form games constitute the set of dyadic rational games,
given by
$\mathbb{D} = \{ m/2^k : m \in \mathbb{Z}, k \in\mathbb{N}_0 = \mathbb N \cup \{0 \}\} $.
\end{defi}

We also refer to dyadic rational games simply as \emph{dyadics}. Note that the set of integer games $\mathbb I $ is a subset of $\mathbb D$.

We now establish some basic properties of dyadic rational games, leading to the determination of their canonical forms and highlighting gaps in precision in the existing literature.


\begin{prop}\label{prop:halfkalt}
 For each $k \in \mathbb{N}_0$, the game $1/2^k > 0$.
\end{prop}

\begin{proof}
We prove using induction on $k \in \mathbb{N}_0$.

Consider $k = 0$. The game is $1/2^0 =1=\cg{0}{\varnothing}$. By Proposition~\ref{prop:Lwinnalt}, we have $1>0$.

Next, by induction, let's assume that the statement is true for all $k = n$. 

Now consider $k=n+1$. By definition, the game $1/2^{n+1} = \cg{0}{1/2^{n}}$.  

If Left starts, she moves to $0$ and wins. If Right starts, he moves to $1/2^{n}$, which is by induction greater than $0$. So Left wins again.

Therefore, Left always wins the game $1/2^{n+1}$, and it follows that $1/2^{n+1} > 0$. 

This completes the proof by induction.
\end{proof}
As an immediate consequence of Proposition~\ref{prop:halfkalt}, we obtain the following corollary.
\begin{cor}\label{cor:dyadposalt}
    For all $m \in \mathbb N$ and $k \in \mathbb N_0$, the game $m/2^k > 0$.
\end{cor}

\begin{proof}
    By definition, the game $m/2^k$ is the disjunctive sum of $m$ copies of $1/2^k$. By Proposition~\ref{prop:halfkalt}, for all $k \in \mathbb{N}_0$, we have $1/2^k > 0$.  Therefore, by repeatedly applying the translation-invariant property of $\mathbb{G}$, we have that for all $m \in \mathbb{N}$ and $k \in \mathbb{N}_0$, $m/2^k > 0$.
\end{proof}
Corollary~\ref{cor:dyadposalt} leads to the following observation.
\begin{obs}\label{obs:dyadnegalt}
For all $m \in \mathbb N$ and $k \in \mathbb N_0$, by Corollary~\ref{cor:dyadposalt}, the game $m/2^k > 0$. Therefore, its inverse satisfies $-m/2^k<0$.
\end{obs}
We next compare dyadic rational games of the form $1/2^k$ and establish a strict ordering based on the exponent.
\begin{prop}\label{prop:dyadcomparealt}
 For all $m , k \in \mathbb N_0$, if $m > k$, then the game
$$1/2^m <1/2^k.$$
\end{prop}

\begin{proof}
   We prove using induction on $k \in \mathbb{N}_0$. 

   Consider the base case $k = 0$. We need to prove that for all $m>0$, 
   $$1/2^m<1/2^0=1.$$
   Consider the game $1/2^m -1$. If Left starts, she can only play to $-1$, which Right wins as $-1<0$. On the other hand, if Right starts, he plays to $1/2^{m-1}-1$. If $m=1$, then Right wins immediately. If $m>1$, Left can only play to $-1$, which again Right wins. Therefore, Right always wins the game $1/2^m-1$, and it follows that $1/2^m<1$.

   Next, by induction, suppose that the proposition is true for $k = n$, that is, for all $m>n$,
   $$1/2^m <1/2^n.$$

   Now consider $k=n+1$. We need to prove that for all $m>n+1$, 
   $$1/2^m<1/2^{n+1}.$$
  Consider the game $1/2^m-1/2^{n+1}$. If Left starts, she plays to either $1/2^m - 1/2^n$ or $-1/2^n$. Right wins $1/2^m - 1/2^n$ by induction and wins $-1/2^n$ as $-1/2^n<0$.

   If Right starts, he plays to $1/2^{m-1} - 1/2^{n+1}$. In response, if Left plays in the first component to $-1/2^{n+1}$, then since the game is less than $0$, Right wins. If she plays in the second component to $1/2^{m-1} - 1/2^n$, then again, by induction Right wins. 

   Therefore, in all cases, Right wins the game $1/2^m - 1/2^{n+1}$, and it follows that $1/2^m < 1/2^{n+1}$.
   
This completes the proof by induction.
\end{proof}

The following result is an important step towards game reduction for any game in $\mathbb{D}$.

\begin{prop}\label{prop:dyadsumalt}
For all $k \in \mathbb{N}$, the sum 
$1/2^k + 1/2^k = 1/2^{k-1}$.
\end{prop}

\begin{proof}
   We prove using induction on $k \in \mathbb{N}$. 

   Consider the base case $k = 1$. We need to prove $$1/2 + 1/2 = 1.$$
   Consider the game $1/2 + 1/2 - 1$. If Left starts, she can only play to $1/2-1$. In response, Right plays to $-1$, which he wins as $-1 < 0$.  On the other hand, if Right starts, he plays to either $1+ 1/2 -1 = 1/2$ or to $1/2+ 1/2$. In both cases, Left wins as the games are greater than $0$.

 Next, by induction, suppose that the proposition is true for $k = n$, that is, $$1/2^n + 1/2^n = 1/2^{n-1}.$$
 Now consider $k=n+1$. We need to prove $$1/2^{n+1} + 1/2^{n+1} = 1/2^{n}.$$ 
 Consider the game $1/2^{n+1} + 1/2^{n+1} - 1/2^{n}$. If Left starts, she plays to either $1/2^{n+1} - 1/2^n$ or to $1/2^{n+1}+ 1/2^{n+1} - 1/2^{n-1}$. By Proposition~\ref{prop:dyadcomparealt}, $1/2^{n+1} - 1/2^n<0$, so Right wins. By induction, the other option $1/2^{n+1}+ 1/2^{n+1} - 1/2^{n-1}$ is equivalent to $1/2^{n+1}+ 1/2^{n+1} - 1/2^n - 1/2^n$. By Proposition~\ref{prop:dyadcomparealt}, the game is negative, so Right wins.

 If Right starts, he plays to either $1/2^n + 1/2^{n+1} - 1/2^n = 1/2^{n+1}$ or to $1/2^{n+1} + 1/2^{n+1}$. In both cases, Left wins as the games are greater than $0$.

   Therefore, $1/2^{n+1} + 1/2^{n+1} - 1/2^{n}=0$, and it follows that
   $1/2^{n+1} + 1/2^{n+1} = 1/2^n$.
   
This completes the proof by induction.
\end{proof}

As in the case of integer games, where an alternative expression is provided by Lemma~\ref{lem:intsimplalt}, we now present an analogous expression for dyadic rational games.

\begin{lem}\label{lem:dyadsimplalt}
For all $m \in \mathbb Z$ and $k \in \mathbb N_0$, $$m/2^k=\cg{(m-1)/2^k}{(m+1)/2^k}.$$
\end{lem}

\begin{proof}
The case when $k=0$ corresponds to integer games. The corresponding result is proved in Lemma~\ref{lem:intsimplalt}. We now prove the result for $k>0$. 

\noindent \textbf{Case 1:}\label{lemprf:dyadsimplaltcase1} $m>0$. From the definition of the sum of games and by using Proposition~$\ref{prop:dyadsumalt}$:
\begin{align*}
 m/2^k &= (m-1)/2^k + 1/2^k \\
    &= \cg{(m-1)/2^k}{(m-1)/2^k + 1/2^{k-1}}\\
    &= \cg{(m-1)/2^k}{(m-1)/2^k+2/2^{k}}\\
    &= \cg{(m-1)/2^k}{(m+1)/2^k}.
\end{align*} 

\noindent \textbf{Case 2:} $m=0$. We need to prove $\cg{-1/2^k}{1/2^k} = 0$. 

If Left starts in the game $\cg{-1/2^k}{1/2^k}$, she plays to $-1/2^k$, which Right wins by Observation~\ref{obs:dyadnegalt}. If Right starts, he plays to $1/2^k$, which Left wins by Corollary~\ref{cor:dyadposalt}. 

Since the second player always wins the game $\cg{-1/2^k}{1/2^k}$, it follows that this game is equal to $0$.

\noindent \textbf{Case 3:} $m<0$. Then $-m>0$, and by Case~\hyperref[lemprf:dyadsimplaltcase1]{1}, 
$$-m/2^k = \cg{(-m-1)/2^k}{(-m+1)/2^k}.$$
This implies 
\begin{align*}
    m/2^k &= -\cg{(-m-1)/2^k}{(-m+1)/2^k}\\
    &= \cg{-(-m+1)/2^k}{-(-m-1)/2^k}\\
    &= \cg{(m-1)/2^k}{(m+1)/2^k}.
\end{align*}
This completes the proof.
\end{proof}

In the literature, to the best of our knowledge, dyadic rational game forms have not been dealt precisely,
 which led to several issues.
For instance, in Siegel’s book~\citep[Theorem 3.6]{S2013}, while identifying the canonical form of dyadic rationals, it is stated that for $n \geq 1$ and odd $m$,
\begin{equation}\label{seigeldyadcanonical}
    m/2^n = \cg{(m-1)/2^n}{(m+1)/2^n}
\end{equation}
is in canonical form. However, this form includes $(m+1)/2^n$ as a Right option, which has a higher birthday than that of $m/2^n$. Thus, the game form in~\eqref{seigeldyadcanonical} must reduce to one with a smaller birthday. Therefore, the game form~\eqref{seigeldyadcanonical} cannot be canonical.

Further, in the proof, Siegel assumes that since $m$ is odd and $n \ge 1$, the game $(m-1)/2^n$ reduces to  $i/2^{n'}$ with $i$ odd and $1 \le n' < n$, and therefore has the same form as in~\eqref{seigeldyadcanonical}. This leads to two issues:
\begin{enumerate}[i)]
    \item  The Left option $(m-1)/2^n$ in~\eqref{seigeldyadcanonical} is also not in canonical form.
    \item The reduction of $(m-1)/2^n$ to a form $i/2^{n'}$ with $i$ odd and $1 \le n' < n$ does not always hold. For example, if $m = 5$ and $n = 1$, then $(m-1)/2^n = 4/2 = 2$, which is even. In the proof, there is no discussion about the canonical form of such subpositions.
\end{enumerate}
Thus, Theorem~\citep [Theorem 3.6]{S2013} is not correctly stated, and its proof is flawed. In the following, we derive the correct canonical forms for dyadic rational games, addressing these inconsistencies. For a game $G \in \mathbb G$, we use $[G]$ to denote its canonical game form.

By Proposition~\ref{prop:dyadsumalt}, any dyadic rational game of the form $m/2^k$ with even $m \in \mathbb{Z}$ and $k \in \mathbb{N}$ reduces either to an integer game form or to a form with an odd numerator. Hence, when $k \in \mathbb{N}$, it suffices to consider only those cases where $m$ is odd.

We are well prepared to find the canonical forms of dyadic games. 
\begin{prop}[Canonical Form Dyadics]\label{prop:simpledyadicscanonical}
For any dyadic:
\begin{enumerate}[(i)]
 \item For any $m \in \mathbb{Z}$, $[m] \cong m $.
    \item For any $k \in \mathbb N$,
    $$[1/2^k]\cong1/2^k.$$
    \item For any $m,k \in \mathbb N$ with odd $m$, 
\begin{equation}\label{eq:positivedyadiccanonical}
    [m/2^k]\cong \cg{[i/2^{k-\alpha}]}{[j/2^{k-\beta}]}
\end{equation}
    where $\alpha, \beta \in \mathbb{N}$ with $k - \alpha \ge 0$ and $k - \beta \ge 0$, and $i, j \in \mathbb{N}_0$ are chosen so that
    $$i/2^{k-\alpha} = (m-1)/2^k \text{ and } j/2^{k-\beta} = (m+1)/2^k,$$
   with $i$ odd if $k-\alpha > 0$ and $j$ odd if $k-\beta > 0$.\footnote{The alternate representation $[m/2^k] \cong \cg{[(m-1)/2^k]}{[(m+1)/2^k]}$ is also valid. However, we prefer the form in~\eqref{eq:positivedyadiccanonical}, as it provides a recursive structure.}
    \item For any $m,k \in \mathbb N$ with odd $m$, $[-m/2^k] \cong -[m/2^k]$
\end{enumerate}
\end{prop}
\begin{proof}
We prove each part separately.\\

\noindent\textit{Proof of (i).}\label{prf:dyadiccanonicalintegers} 
By definition, the zero game is given by $0 \cong \cg{\varnothing}{\varnothing}$. Since it has no Left or Right options, it is already in canonical form. Therefore, $[0] \cong 0$.




Let $m \in \mathbb N$. By definition $m:=\cg{m-1}{\varnothing}$. By induction, assume that $m-1$ is in canonical form. In the game $m$, 
 Left has exactly one option and Right has none, so there is no domination. The Left option $m-1$ has no Right responses, and hence it is not reversible. There are no Right options to check for reversibility. Thus, $m$ has no dominated or reversible options. Further, since by induction all its subgames are in canonical form, $m$ is in canonical form.  Therefore, $[m] \cong m$. Hence, its inverse $-m$ is also in canonical form. Therefore, $[-m] \cong -m$.\\



\noindent\textit{Proof of (ii).}\label{prf:dyadiccanonicaldyadic} Recall that $1/2^0 := 1$.
 Let $k \in \mathbb N$. By definition $1/2^k:= \cg{0}{1/2^{k-1}}$. 
 %
 %
By induction, assume that $1/2^{k-1}$ is in canonical form. In the game $1/2^k$, both players have only one option, so there is no domination. The Left option $0$ has no Right response, so it is not reversible. The Right option $1/2^{k-1}$ has a Left response $0$, but since $0 \not\ge \cg{0}{1/2^{k-1}}$, it is not reversible. Hence, $1/2^k$ has no dominated or reversible options. By induction, all of its options are in canonical form. Therefore, $1/2^k$ is in canonical form, and hence $[1/2^k] \cong 1/2^k$.\\

\noindent\textit{Proof of (iii).}\label{prf:dyadiccanonicalsum}  Consider
\begin{equation}
    G \cong \cg{[i/2^{k-\alpha}]}{[j/2^{k-\beta}]},\label{prf:dyadiccanonicaleq1}
\end{equation}
where $\alpha, \beta, i, j$ are as in the statement.


     From Lemma~\ref{lem:dyadsimplalt}, we have $G = m/2^k$. It only remains to show that $G$ is in canonical form. We prove this by induction on odd integers $m$.

     For the base case $m=1$, we have 
     $$i/2^{k-\alpha} = (m-1)/2^k = 0 \text{ and } j/2^{k-\beta} = (m+1)/2^k = 2/2^k = 1/2^{k-1},$$ 
     so
     $$G = \cg{[0]}{[1/2^{k-1}]}.$$
By part~\hyperref[prf:dyadiccanonicaldyadic]{(ii)}, $G = \cg{0}{1/2^{k-1}}$ is in canonical form.

       Next, assume by induction that for odd all $m\le n$, where $n$ is odd, the game $G$ defined by~\eqref{prf:dyadiccanonicaleq1} is in canonical form. Consider $m=n+2$. Then
       $$G = \cg{[i/2^{k-\alpha}]}{[j/2^{k-\beta}]},$$
       with
        $$i/2^{k-\alpha} = (n+1)/2^k \text{ and } j/2^{k-\beta} = (n+3)/2^k.$$

       In $G$, both players have exactly one option, so there is no dominating option. To check reversibility, consider the Left option $[i/2^{k-\alpha}]$. By induction, its Right option is 
       $$[i_R/2^{k-\alpha-\beta_1}],$$ 
       where $ \beta_1 \in \mathbb{N}$ with $k-\alpha -\beta_1\ge 0$ and $i_R \in \mathbb N_0$ such that $i_R/2^{k-\alpha-\beta_1} = (i+1)/2^{k-\alpha}$ with $i_R$ odd if $k-\alpha -\beta_1> 0$. Thus, we have
       \begin{align*}
           i_R/2^{k-\alpha-\beta_1} &= (i+1)/2^{k-\alpha}\\
           &= i/2^{k-\alpha} + 1/2^{k-\alpha}\\
           &= (m-1)/2^k +  1/2^{k-\alpha}\\
           &= (m-1+2^\alpha)/2^k.
       \end{align*}
       Since $\alpha \in \mathbb N$, we have $m-1+2^\alpha\ge m$, so
       $i_R/2^{k-\alpha-\beta_1} \not \le m/2^k =G$.
    Thus, the Left option is not reversible.

       By a similar analysis, we get that the Right option in $G$ is not reversible.
       
       Therefore, $G$ has no dominated or reversible options. Since, by induction, all its options are in canonical form, it follows that $G$ is in canonical form.\\

       \noindent\textit{Proof of (iv).} Since the negative of a canonical form is also in canonical form. Thus, the result follows from part~\hyperref[prf:dyadiccanonicalsum]{(iii)}.
\end{proof}

For any $k \in \mathbb{N}$, the game form $2/2^k$ belongs to the set of dyadic rational games $\mathbb{D}$. However, as shown in Proposition~\ref{prop:simpledyadicscanonical}, this form is not in canonical form, since it reduces to $1/2^{k-1}$. Therefore, not all game forms in $\mathbb{D}$ are in canonical form. To address this, we define a new set $\mathbb{D}_{\text{c}}$ consisting only of the canonical forms of elements in $\mathbb{D}$: 
$$\mathbb{D}_\text{c} = \{[G] \,|\, G\in \mathbb G \text{ and } G=x \text{ for some } x\in \mathbb D  \}.$$
Consider the game $1/2$. By definition, $1/2= \cg{0}{1}$, which is also in canonical form. Therefore, 
$$\cg{0}{1} \in \mathbb D \cap \mathbb{D}_{\text{c}}.$$
Next, consider the game $3/2$. By definition, it is the disjunctive sum $1/2 + 1/2 +1/2$,  and hence its game form is determined as follows:
    \begin{align}
    3/2 &\cong 1/2 + 1/2 +1/2 \nonumber\\
    &\cong \cg{0}{1}+\cg{0}{1}+\cg{0}{1} \nonumber\\
    &\cong \cg{0}{1} + \cg{\cg{0}{1}}{1+\cg{0}{1}} \nonumber\\
          & \cong {\big\{}
        \cg{\cg{0}{1}}{1 + \cg{0}{1}}, \, 
        \cg{0}{1} + \cg{0}{1} {\big|} \begin{aligned}[t]
        & \,1 + \cg{\cg{0}{1}}{1 + \cg{0}{1}}, \\
        &\, 1 + \cg{0}{1} + \cg{0}{1} {\big \}}
    \end{aligned} 
   \label{ex:3/2literal}
    \end{align}  
    
Thus, the game form as in~\eqref{ex:3/2literal} belongs to $\mathbb D$. However, the canonical form of $3/2$, as given by Proposition~\ref{prop:simpledyadicscanonical}, is $\cg{1}{2}$. This implies:
\begin{enumerate}[i)]
    \item The game form in~\eqref{ex:3/2literal} is in $\mathbb D$ but not in $\mathbb{D}_{\text{c}} $, as it is not canonical.
    \item The game form $\cg{1}{2}$ is in $\mathbb{D}_{\text{c}}$ but not in $\mathbb D$,  since it cannot be obtained using only the definition of dyadics, without applying reduction.
    \item $\mathbb{D}_{\text{c}} $ is not a group, as it is not closed under disjunctive sum.
\end{enumerate}

This example highlights that while $\mathbb{D}_{\text{c}}$ contains only canonical forms, the set $\mathbb{D}$, fails to include some canonical game forms. In fact, neither $\mathbb D  \subseteq \mathbb{D}_{\text{c}}$ nor  $\mathbb{D}_{\text{c}} \subseteq\mathbb D $ holds. Therefore,  to work within a group theoretic framework, we now introduce the group generated by $\mathbb{D}_{\text{c}}$.

We define $\langle \,\mathbb{D}_{\text{c}} \rangle$ to be the subgroup of $\mathbb{G}$ generated by $\mathbb{D}_{\text{c}}$. By definition of dyadics, we have $\mathbb{D} \subset \langle \,\mathbb{D}_{\text{c}} \rangle$. However, there is an additional subtlety: the literal game form $\cg{\cgstar}{\cgstar} \notin \langle\, \mathbb{D}_{\text{c}} \rangle$, even though its game value $0$ lies in $\mathbb{D}_{\text{c}}$, and hence in $\langle \,\mathbb{D}_{\text{c}} \rangle$.

Therefore, to rigorously capture the collection of all games whose game values lie in $\mathbb{D}_{\text{c}}$, we define:
$$\widetilde{\mathbb D} = \{G\in \mathbb G \;|\; G=x \text{ for some } x \in \mathbb{D}_{\text{c}}\}.$$
Now, by definition of $\mathbb{D}_{\text{c}}$, for every $x \in \mathbb{D}_{\text{c}}$, there exists a game $y \in \mathbb D$ such that $x=y$. Hence, we can equivalently write:
$$\widetilde{\mathbb D} = \{G\in \mathbb G \;|\; G=x \text{ for some } x \in \mathbb{D}\}.$$
We now prove that $\widetilde{\mathbb D}$ forms a subgroup of $\mathbb G$.

\begin{thm}\label{thm:tildeDgroup}
    The set $\widetilde{\mathbb D}$  is a subgroup of $\mathbb G$.
\end{thm}

\begin{proof}
    To show that $\widetilde{\mathbb D} \subset \mathbb G$,  is a subgroup, it suffices to prove that for any $G, H \in \widetilde{\mathbb D}$, the game $G-H \in \widetilde{\mathbb D}$.

    Since $G, H \in \widetilde{\mathbb{D}}$, by definition there exist $x, y \in \mathbb{D}$ such that $G = x$ and $H = y$. Without loss of generality, for some $n, m \in \mathbb{Z}$ and $j, k \in \mathbb{N}_0$ with $k \geq j$ let $x= n/2^k$ and $y = m/2^{k-j}$. Then by using Proposition~\ref{prop:dyadsumalt}, we get 
    $$y = 2^jm/2^k,$$ 
    and hence, 
    $$G - H = x-y = n/2^k - 2^jm/2^k = (n- 2^jm)/2^k.$$
    Now since, $(n- 2^jm)/2^k \in \mathbb D$, it follows that $G - H \in \widetilde{\mathbb{D}}$.
\end{proof}

Now, since $\mathbb{D}_{\text{c}} \subset \widetilde{\mathbb D} $ and $\widetilde{\mathbb D}$ is a group, therefore, $\langle \,\mathbb{D}_{\text{c}} \rangle \subset \widetilde{\mathbb D}$.

The following Venn diagram illustrates the relationship between the sets $\mathbb{D}$, $\mathbb{D}_{\text{c}}$, $\langle \mathbb{D}_{\text{c}} \rangle$, and $\widetilde{\mathbb{D}}$.

\begin{figure}[ht]
    \centering
\begin{tikzpicture}[scale=1.4,thick]
  \draw[rounded corners=10pt] (0,0) rectangle (5,3);
  \node at (4.7,2.7) {$\widetilde{\mathbb{D}}$};

  \draw[rotate=0] (2.5,1.5) ellipse (2 and 1.3);
  \node at (2.5,2.5) {$\langle\,\mathbb{D}_{\text{c}}\rangle$};

  \draw (2,1.5) circle (0.8);
  \node at (1.8,1.5) {$\mathbb{D}$};

  \draw (3,1.5) circle (0.8);
  \node at (3.2,1.5) {$\mathbb{D}_{\text{c}}$};

\end{tikzpicture}
\caption{Venn diagram illustrating the relations between $\mathbb{D}$, $\mathbb{D}_{\text{c}}$, $\langle\, \mathbb{D}_{\text{c}} \rangle$, and $\widetilde{\mathbb{D}}$.}
    \label{fig:venn-dyadic}
\end{figure}
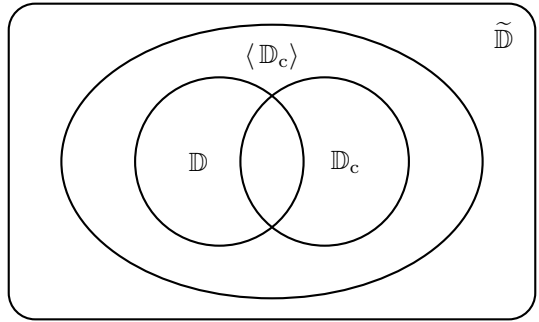

Now, to understand why these distinctions are important, consider the game $G = \cg{\cgstar}{\cgstar}$. Observe that, it does not hold that, 
\begin{equation}\label{eq:zugzwang}
   \text{for every } G^L \in G^\mathcal{L} \text{ and every } G^R \in G^\mathcal{R},  G^L < G < G^R.
\end{equation}
However, its game value is $0$, which does satisfy this property. In the literature, the property~\eqref{eq:zugzwang} has often been used under the name of dyadics (see e.g.~\citep[Theorem 3.10]{S2013}), but as this example shows, it does not hold for every game that equals a dyadic. On the other hand, from Proposition~\ref{prop:simpledyadicscanonical}, it is easy to check that if a game $G$ belongs to $\mathbb{D}_{\text{c}}$, then it does satisfy property~\eqref{eq:zugzwang}. We will see that the group $\langle\, \mathbb{D}_{\text{c}} \rangle$ also inherits property~\eqref{eq:zugzwang}. The relevance of this property will become more apparent in the next section.




We conclude this section by proving some additional structural properties of the set $\widetilde{\mathbb D}$. We start by showing it is a \textbf{totally ordered group}.

\begin{prop}\label{prop:tildeDtotalorder}
    The set $\widetilde{\mathbb D}$ is a totally ordered group under the game relation ``$\ge$''.
\end{prop}

\begin{proof}
By Theorem~\ref{thm:tildeDgroup}, the set $\widetilde{\mathbb D}$ is a subgroup of $\mathbb{G}$. Since the game relation ``$\ge$'' is translation-invariant, to show that $\widetilde{\mathbb{D}}$ is a totally ordered group, it suffices to prove that the relation ``$\ge$'' is a total order on $\widetilde{\mathbb D}$.

Let $G, H \in \widetilde{\mathbb{D}}$. By definition, there exist dyadic rationals $x, y \in \mathbb{D}$ such that $G = x$ and $H = y$. Without loss of generality, suppose that for some $n, m \in \mathbb{Z}$ and $j, k \in \mathbb{N}_0$ with $k \geq j$, we have
$$x= n/2^k \text{ and } y = m/2^{k-j}.$$
Then, by Proposition~\ref{prop:dyadsumalt}, we can write 
$$y = 2^j m/2^k.$$
Since $\mathbb{Z}$ is totally ordered, we have either $n \ge 2^j m$ or $2^j m \ge n$. This implies 
$$\text{either }\quad  n/2^k \ge 2^j m/2^k \quad \text{or} \quad  2^j m/2^k \ge n/2^k,$$
that is, either $x \ge y$ or $y \ge x$, and hence $G \ge H$ or $H \ge G$.

This shows that the relation ``$\ge$'' is total on $\widetilde{\mathbb D}$, completing the proof.
\end{proof}
By Proposition~\ref{prop:tildeDtotalorder}, we have that ``$\ge$'' is a total order on $\widetilde{\mathbb D}$. Consequently, any subset of $\widetilde{\mathbb D}$ inherits this total order under the same relation. In particular, the sets $\mathbb{D}$, $\mathbb{D}_{\text{c}}$, and $\langle\, \mathbb{D}_{\text{c}} \rangle$ are each totally ordered with respect to ``$\ge$''. This result also implies that no game in $\widetilde{\mathbb D}$ belongs to the outcome class $\nN$.


We now show that the set  $\widetilde{\mathbb D}$ is \textbf{dense-in-itself}.

\begin{lem}\label{lem:dyadtildedensealt}
    Consider $G, H \in \widetilde{\mathbb D}$ such that $G<H$. Then there exists $K \in \widetilde{\mathbb D}$ such that $G<K<H$.
\end{lem}

\begin{proof}

Let $G, H \in \widetilde{\mathbb{D}}$. By definition, there exist $x, y \in \mathbb{D}$ such that $G = x$ and $H = y$. Without loss of generality, for some $n, m \in \mathbb{Z}$ and $j, k \in \mathbb{N}_0$ with $k \geq j$ let $x= n/2^k$ and $y = m/2^{k-j}$ such that $n/2^k < m/2^{k-j}$. Since $n/2^k < m/2^{k-j}$, it follows that $n < 2^j m$.

    Define a dyadic rational game $K$  by
   $$K = (n+2^jm)/2^{k+1}.$$
  We now show that $G<K<H$.    First, observe,
   \begin{align*}
   G-K &= x-K \\
   &=n/2^k -  (n+2^jm)/2^{k+1}\\
       &=2n/2^{k+1} - (n+2^jm)/2^{k+1}\\
       &= (2n-n-2^jm)/2^{k+1}\\
       &= (n-2^jm)/2^{k+1}<0
       \end{align*}
       since $n < 2^j m$. Hence, $G < K$.
       
       Next, consider, 
   \begin{align*}
   K-H &=K-y\\
&=  (n+2^jm)/2^{k+1} - m/2^{k-j}\\
       &= (n+2^jm)/2^{k+1} - 2^{j+1}m/2^{k+1}\\
       &= (n+2^jm - 2^{j+1}m)/2^{k+1}\\
       &= (n-2^jm)/2^{k+1}<0,
       \end{align*}
     again using $n < 2^j m$. Hence, $K < H$.

      Therefore, we have $G < K < H$.
\end{proof}
From Lemma~\ref{lem:dyadtildedensealt}, we make the following immediate observation.
\begin{obs}\label{obs:dyaddensealt}
    By a proof similar to that of Lemma~\ref{lem:dyadtildedensealt}, we see that both $\mathbb{D}$ and $\mathbb{D}_{\text{c}}$ are dense-in-itself.
\end{obs}

\section{Numbers and zugzwangs}\label{sec:numbersalt}

In combinatorial game theory, \textit{numbers} represent a fundamental class of game values. They were first introduced and systematically studied by Conway in his work \textit{On Numbers and Games}~\citep{ONAG}, with the first edition published in 1976. He observed that a certain class of games, that exhibit real number-like properties, arises naturally from the analysis of combinatorial games.

Conway defined \textit{numbers} constructively, as follows: 

\begin{defi}[Conway Number~\citep{ONAG}]\label{thm:numberconway}
     A game $G = \cg{G^\mathcal{L}}{G^\mathcal{R}}$ is a Conway number if, for every $G^L \in G^\mathcal{L}$ and every $G^R \in G^\mathcal{R}$, we have $G^L  \cglfuz G^R$, and each option is a Conway number.
\end{defi}

For brevity, we refer to a Conway number as a \textit{C-number}. These are also known as \textit{surreal numbers}.

By Conway’s definition, $0 = \cg{\varnothing}{\varnothing}$ is the first constructed C-number. But the game $\cgstar = \cg{0}{0}$ is not a C-number, even though its options are C-numbers, because its Left and Right options are equal. Since $\cg{\cgstar}{\cgstar} = 0$, this highlights a subtle inconsistency in the definition: a game that is equal to a C-number may itself fail to satisfy the definition of a C-number.

In an alternative approach, Siegel introduced a different definition of numbers in his book~\citep{S2013}: 
\begin{defi}[Siegel Number~\citep{S2013}]\label{thm:numbersiegel}
     A game $G = \cg{G^\mathcal{L}}{G^\mathcal{R}}$ is a Siegel number if, for every 
     $G^L \in G^\mathcal{L}$ and every $G^R \in G^\mathcal{R}$, we have $G^L < G^R$, and each option is a Siegel number.
\end{defi}

For brevity, we refer to a Siegel number as an \textit{S-number}.

The same issue remains: the game $\cg{\cgstar}{\cgstar}$ is still not an S-number, while the game $0$ is. Both these definitions are constructive. If we strictly follow their constructive definitions, such inconsistencies do not arise. This suggests that the definitions implicitly assume the use of canonical forms of games, even though this is not stated explicitly. As a result, when applied to the literal form of a game, the definitions show inconsistency. We will see in the following that if we restrict to canonical forms of games, then both definitions exactly characterize the class of dyadic rational games.


But before that, in view of \eqref{eq:zugzwang}, we introduce another important notion, called \textit{zugzwangs}; these are games in which no player has an incentive to start. Such situations occur in the well-known combinatorial game of  \textsc{Chess}~\citep{zugzwangChess}. 

For example, consider the game $3/2 = \cg{5/4}{7/4}$. If Left starts, she plays to $5/4$, which is strictly less than $3/2$, thereby losing her advantage. If Right starts, he plays to $7/4$, which is strictly greater than $3/2$, thus giving an advantage to Left. 



\begin{defi}[Zugzwang]\label{def:zugzwangalt}
    A game $G = \cg{G^\mathcal{L}}{G^\mathcal{R}} \in \mathbb G$ is a zugzwang if, for every $G^L \in G^\mathcal{L}$ and every $G^R \in G^\mathcal{R}$, we have $G^L < G < G^R$, and each option is a zugzwang. We denote 
$\zug = \{G \in \mathbb G \mid G \text{ is a zugzwang}\}$.
\end{defi}

Observe that this definition of zugzwang is a stronger notion, as it also requires that all subpositions of the game be zugzwangs.

By definition, every zugzwang is an S-number. We next prove the converse, that every S-number is a zugzwang.

\begin{lem}\label{lem:seigelzugzwang}
Consider a game  $G \in \mathbb G$. If $G$ is an S-number then $G$ is also a zugzwang.
\end{lem}

\begin{proof} Let $G \in \mathbb G$ be an S-number. Then, for every 
     $G^L \in G^\mathcal{L}$ and every $G^R \in G^\mathcal{R}$, we have $G^L < G^R$. For any $G^L \in G^\mathcal{L}$, consider the game $G-G^L$. 
    
    If Left starts, she plays to $G^L -G^L = 0$ and wins. 
    
    If Right starts, he plays to either $G^R - G^L$ or $G-G^{LL}$. In the first case, since $G^L < G^R$ by the definition of an S-number, Left wins $G^R - G^L$ playing first. In the second case, Left plays to $G^L - G^{LL}$ and wins by induction.
    
    Therefore, Left wins regardless of who starts. Hence, $G - G^L>0$, which implies $G^L<G$.

A similar analysis shows that $G < G^R$. Thus, if $G$ is an S-number, it satisfies $G^L < G < G^R$. Furthermore, since every option of $G$ is an S-number, this implies that each option also satisfies this property. Hence, $G$ is a zugzwang.
\end{proof}

By definition, every zugzwang is a C-number. A similar proof to that used for S-numbers shows that if $G$ is a C-number, then $G$ is also a zugzwang. Hence, 
\begin{align*}
    \{G \in \mathbb G \mid G \text{ is a C-number}\} &= \{G \in \mathbb G \mid G \text{ is a zugzwang}\}\\
    &= \{G \in \mathbb G \mid G \text{ is an S-number}\}.
\end{align*}
We now prove that the set of zugzwangs forms a group.
 \begin{thm}\label{thm:zugzwanggroupalt}
     The set of zugzwangs is a subgroup of $\mathbb G$.
 \end{thm}

 \begin{proof}
     Let $G, H$ be two zugzwangs. Consider their difference:
     $$G-H = \cg{G^{\mathcal L} - H, G-H^{\mathcal R}}{G^{\mathcal R} - H, G-H^{\mathcal L}}.$$
     It is sufficient to show that $G-H$ is a zugzwang. 
     
     Since $G$ is a zugzwang, for every $G^L \in G^\mathcal{L}$ and every $G^R \in G^\mathcal{R}$, we have $G^L < G < G^R$. Similarly, since $H$ is a zugzwang, for every $H^L \in H^\mathcal{L}$ and every $H^R \in H^\mathcal{R}$, we have $H^L < H < H^R$. Therefore, we obtain the following inequalities:
     \begin{enumerate}[i)]
         \item for every $G^L \in G^\mathcal{L}$, we have $G^L -H < G-H$,
         \item for every $G^R \in G^\mathcal{R}$, we have $G-H<G^R-H$,
         \item for every $H^L \in H^\mathcal{L}$, we have $G-H<G-H^L$,
         \item for every $H^R \in H^\mathcal{R}$, we have  $G-H^R<G-H$.
     \end{enumerate}
     Further, since the options of $G$ and $H$ are also zugzwangs, it follows by induction that all the options of $G - H$, namely those games in   $G^{\mathcal L} - H, G-H^{\mathcal R}, G^{\mathcal R} - H$ and $ G-H^{\mathcal L}$,  are zugzwangs as well.
     
     This completes the proof.
 \end{proof}

 From Proposition~\ref{prop:simpledyadicscanonical}, we observe that all games in $\mathbb{D}_{\text{c}}$ are zugzwangs. Since the set of zugzwangs forms a group, it follows that $\langle \, \mathbb{D}_{\text{c}} \rangle$ is a subgroup of the set of zugzwangs. Next, we make two key observations:
\begin{enumerate}[i)]
    \item The game form $\cg{n-1}{n+1}$ is a zugzwang, but it cannot be represented as sum of games  only from $\mathbb{D}_{\text{c}}$ without applying reductions. Therefore, we have strict inclusion: 
    $$ \langle\, \mathbb{D}_{\text{c}} \rangle \subset \zug.$$
    \item The game $\cg{\cgstar}{\cgstar}$ is not a zugzwang, while its canonical form is $0$, which lies in $\mathbb D$. Hence, 
    $$ \widetilde{\mathbb D} \not \subset \zug .$$
\end{enumerate}

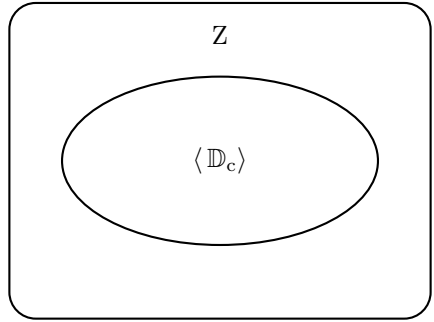
\begin{figure}[ht]
    \centering
\begin{tikzpicture}[scale=1.4,thick]
  \draw[rounded corners=10pt] (0,0) rectangle (4,3);
  \node at (2,2.7) {$\zug $};


  \draw[rotate=0] (2,1.5) ellipse (1.5 and 0.8);
  \node at (2,1.5) {$\langle\,\mathbb{D}_{\text{c}}\rangle$};



\end{tikzpicture}
\caption{This Venn diagram illustrates the relations between the groups $\langle\, \mathbb{D}_{\text{c}} \rangle$ and zugzwangs.}
    \label{fig:vennzugzwang}
\end{figure}

Therefore, we have the Venn diagram in Figure~\ref{fig:vennzugzwang}, which illustrates the relationship between the sets $ \langle\, \mathbb{D}_{\text{c}} \rangle$ and $ \zug $.

This raises natural questions:
\begin{enumerate}[i)]
    \item Is
$ \;\zug \subset \widetilde{\mathbb D}$?
\item How can we completely characterize $\widetilde{\mathbb{D}}$, that is, what property must a literal form game satisfy to be equal to a dyadic?
\end{enumerate}



Recall from Lemma~\ref{lem:GLGGR} that for any game $G \in \mathbb{G}$, we have $G^L \cglfuz G \cglfuz G^R$. Motivated by this, and toward answering the above question, we now define the notion of a \textit{fitting set}.

\begin{defi}[Fitting Set]
    For a game $G \in \mathbb G$, its fitting set is given by:
    $$\mathcal{F}(G) = \{x \mid x\in \mathbb{D} \text{ and,  for every } G^L \text{ and every } G^R,  G^L \cglfuz x \cglfuz G^R\}.$$
\end{defi}

Let us now understand the fitting set through some examples. We will see examples where the fitting set is empty, contains a single element, forms an infinite set or contains everything.

\begin{ex}\label{ex:starfittingset}
    Let $G= \cgstar = \cg{0}{0}$. Then the fitting set is given by
    $$\mathcal F(\cgstar) = \{x \mid x\in \mathbb{D} \text{ and } 0 \cglfuz x \cglfuz 0\}.$$
    Since the dyadics are totally ordered, there is no $x \in \mathbb{D}$ such that $0 < x < 0$. Therefore, 
    $$\mathcal F(\cgstar) = \emptyset.$$
\end{ex}

\begin{ex}\label{ex:starslashstar}
    Let $G  = \cg{\cgstar}{\cgstar}$. Then the fitting set is given by
    $$\mathcal F(\cgstar) = \{ x \mid x\in \mathbb{D} \text{ and } \cgstar\cglfuz x \cglfuz \cgstar\}.$$
    Since $\cgstar\cgfuzzy 0$ and dyadics are totally ordered, it follows that the only $x \in \mathbb{D}$ satisfying $\cgstar \cglfuz x \cglfuz \cgstar$ is $x = 0$. Therefore,
    $$\mathcal F(\cg{\cgstar}{\cgstar}) = \{0\}.$$
\end{ex}

\begin{ex}\label{ex:half}
  Let $G = \cg{1}{2}$. Then the fitting set is given by
       $$ \mathcal{F}(\cg{1}{2}) = \{x \mid x\in \mathbb{D} \text{ and } 1 \cglfuz x \cglfuz 2\}.$$
    Since both $1$ and $2$ are dyadic rationals and $\mathbb{D}$ is totally ordered, this simplifies to
    $$\mathcal{F}(\cg{1}{2}) = \{x \mid x\in \mathbb{D} \text{ and } 1 < x < 2\}.$$
\end{ex}

\begin{ex}\label{ex:zero}
  Let $G = 0 = \cg{\varnothing}{\varnothing}$. Then the fitting set is given by
       $$ \mathcal{F}(\cg{\varnothing}{\varnothing}) = 
\{x \mid x\in \mathbb{D} \} = \mathbb D.$$
\end{ex}

If $\mathcal{F}(G) \neq \emptyset$, we will now derive some properties of the fitting set $\mathcal{F}(G)$. If a game $G$ has the smallest birthday among all its equivalents, we say it is in its \emph{simplest form}.  

\begin{lem}\label{lem:fittingunique}
For any game $G$, if $\mathcal F(G) \neq \emptyset$, then there exists a unique simplest element $x \in \mathcal F(G)$.
\end{lem}

\begin{proof} We proceed by contradiction. Consider a game $G \in \mathbb G$. Suppose, on the contrary, that there exist two distinct elements $x, y \in \mathcal F(G)$ of smallest birthday. Since $\mathcal F(G) \subseteq \mathbb D$, and $\mathbb D$ is totally ordered,  without loss of generality, assume that $x>y$. We prove that there exists a game in $\mathcal F(G)$ of strictly smaller birthday. 

Since $x>y$, Left must have a winning move in the game $x-y$. Therefore, either $x^L \ge y$ or $x \ge y^R$. Now, since $x \in \mathbb D$ is a zugzwang, we have $x>x^L$. Therefore if $x^L \ge y$, then  for every $G^L$ and every $G^R$, we get $G^L\cglfuz y \le x^L<x \cglfuz G^R$. Hence, $x^L \in \mathcal F(G)$, and $x^L$ has strictly smaller birthday than $x$ and $y$. Similarly, if $x \ge y^R$, then  for every $G^L$ and $G^R$, we have $G^R \cggfuz x \ge y^R>y \cggfuz G^L$. Therefore, $y^R \in \mathcal F(G)$, which again has a strictly smaller birthday than $x$ and $y$. This contradicts the assumption that both $x$ and $y$ have the smallest birthday in $\mathcal F(G)$.

Therefore, there cannot be two elements of the smallest birthday in $\mathcal F(G)$.
\end{proof}
We now present one of the most celebrated results in combinatorial game theory, the Simplicity Theorem. It shows that, for a game to be equal to a dyadic, it is enough that $\mathcal F(G) \neq \emptyset$. This seemingly modest condition is, in fact, a subtle and nontrivial requirement, as seen in Example~\ref{ex:starfittingset}.

\begin{thm}[Simplicity Theorem]\label{thm:dyadsimplicityalt} For any game $G \in \mathbb G$, if $\mathcal F(G) \neq \emptyset$,  then $G$ is equal to the unique simplest dyadic $x \in \mathcal{F}(G)$.
\end{thm}
\begin{proof}
    Since $\mathcal F(G)$ is nonempty, by Lemma~\ref{lem:fittingunique}, there exists a unique game of smallest birthday in $\mathcal F(G)$, say $x$. We will show that $G = x$.

    Consider the game $G-x$. 

    If Left starts, she plays to either $G^L-x$ or to $G-x^R$. Since $x \in \mathcal F(G)$, we have $G^L\cglfuz x$, so Right wins $G^L-x$ playing first. Now, since $x$ is of smallest birthday in $\mathcal F(G)$, it follows that $x^R \notin \mathcal F(G)$. Therefore, either there exists a Left option $G^L$ in $G$ such that $x^R \le G^L$ or there exists a Right option $G^R$ in $G$ such that $G^R \le x^R$. 

    If $x^R \le G^L$, then $x<x^R\le G^L$, which contradicts $x\in \mathcal F(G)$. Therefore, there must exist $G^R$ such that $G^R \le x^R$. This implies there is $G^R$ such that $G^R- x^R \le 0$, that is, in the game $G-x^R$, Right plays to $G^R-x^R$ and wins.

    If Right starts, then by an analogous argument, Left wins. 

    Thus, $G-x = 0$, and hence $G=x$.
\end{proof}

Now we will see some applications of the Simplicity Theorem.
\begin{ex}
     Let $G  = \cg{\cgstar}{\cgstar}$.  As found in Example~\ref{ex:starslashstar}, its fitting set is
    $$\mathcal F(\cg{\cgstar}{\cgstar}) = \{0\}.$$
    Since $\mathcal F(\cg{\cgstar}{\cgstar})$ contains only one element, it follows from the Simplicity Theorem~\ref{thm:dyadsimplicityalt} that $\cg{\cgstar}{\cgstar} = 0$.
\end{ex}

\begin{ex}
    Let $G = \cg{1}{2}$.  As found in Example~\ref{ex:half},
 its fitting set is
    $$\mathcal{F}(\cg{1}{2}) = \{x \mid x\in \mathbb{D} \text{ and } 1 < x < 2\}.$$
    Since $3/2$ is the element of smallest birthday in $\mathcal{F}(\cg{1}{2})$, it follows from the Simplicity Theorem~\ref{thm:dyadsimplicityalt} that $\cg{1}{2} = 3/2$.
\end{ex}

The Simplicity Theorem~\ref{thm:dyadsimplicityalt} requires that $\mathcal{F}(G) \neq \emptyset$. For a zugzwang game $G$, we can ensure this condition holds.

\begin{lem} \label{lem:zugzwangfittingset}
If $G \in \mathbb G$ is a zugzwang, then $\mathcal{F}(G) \neq \emptyset$.    
\end{lem}

\begin{proof}
Consider a zugzwang game $G = \cg{G^\mathcal{L}}{G^\mathcal{R}} \in \mathbb G$. By definition, for every $G^L \in G^\mathcal{L}$ and every $G^R \in G^\mathcal{R}$, we have $G^L < G < G^R$, and moreover, each option is also a zugzwang. We need to show that $\mathcal{F}(G) \neq \emptyset$. We prove this by induction on the birthday of game $G$. 

    Consider the base case: b(G) = 0. The only game of birthday $0$ is $0 = \cg{\varnothing}{\varnothing}$. Vacuously,  $\mathcal{F}(0)  = \mathbb D \neq \emptyset$.

    Let us assume the theorem is true for $b(G) \le n$.

    Consider $b(G) = n+1$. In this case, every $G^L \in G^\mathcal{L}$ and $G^R \in G^\mathcal{R}$ has birthday at most $n$. Since they are also zugzwang, by induction, each of them has a nonempty fitting set. Hence, by the  Simplicity Theorem~\ref{thm:dyadsimplicityalt}, they are in $\mathbb D$. 
    
    Since  $\mathbb{D}$ is totally ordered, among the Left options in $G^\mathcal{L}$ there exists an element $G^L$ that dominates all others, and similarly, among the Right options in $G^\mathcal{R}$ there exists an element $G^R$ that dominates all others. Since $G$ is a zugzwang game, we have $G^L < G^R$. Then, by Observation~\ref{obs:dyaddensealt}, there exists a dyadic $x$ such that $G^L < x < G^R$. Hence, $x \in \mathcal{F}(G)$, and therefore $\mathcal{F}(G) \neq \emptyset$.
\end{proof}
We now show that every zugzwang game is equal to a dyadic rational.
\begin{prop} \label{prop:zugzwangequalD}
    If $G \in \mathbb G$ is a zugzwang, then $G \in \widetilde{\mathbb D}$.
\end{prop}

\begin{proof}
Let $G = \cg{G^\mathcal{L}}{G^\mathcal{R}} \in \mathbb G$ be a zugzwang game. By Lemma~\ref{lem:zugzwangfittingset}, we have $\mathcal{F}(G) \neq \emptyset$. Then, by  Simplicity Theorem~\ref{thm:dyadsimplicityalt}, there exists a simplest element $x \in \mathcal{F}(G) \subseteq \mathbb D$ such that $G = x$. Hence, $G \in \widetilde{\mathbb D}$.
\end{proof}

Therefore, by Proposition~\ref{prop:zugzwangequalD}, we have 
$$\{G \in \mathbb G \mid G \text{ is a zugzwang}\} \subseteq \widetilde{\mathbb D}.$$  

Thus, 
$$\{G \in \mathbb G \mid G \text{ is a C-number}\} \subseteq \widetilde{\mathbb D}.$$ 
as well as
$$\{G \in \mathbb G \mid G \text{ is an S-number}\} \subseteq \widetilde{\mathbb D}.$$ 

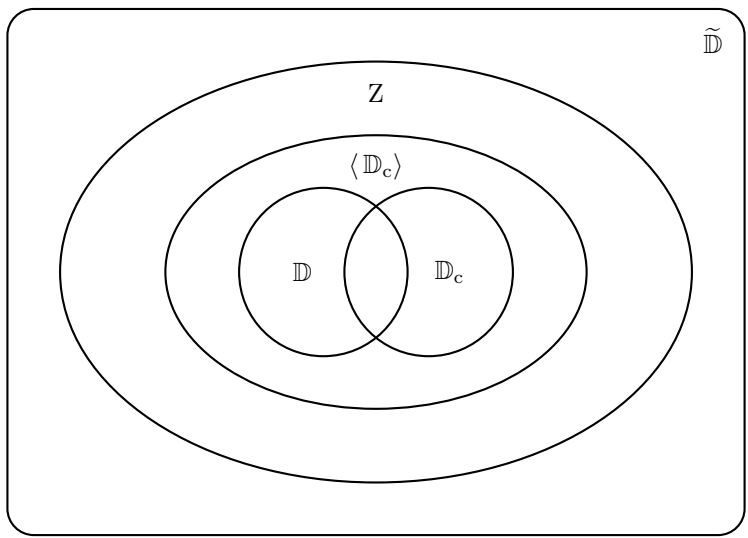
\begin{figure}[ht]
    \centering
\begin{tikzpicture}[scale=1.4,thick]
  \draw[rounded corners=10pt] (0,0) rectangle (7,5);
  \node at (6.7,4.7) {$\widetilde{\mathbb{D}}$};

  \draw[rotate=0] (3.5,2.5) ellipse (3 and 2);
  \node at (3.5,4.2) {$\zug$};

  \draw[rotate=0] (3.5,2.5) ellipse (2 and 1.3);
  \node at (3.5,3.5) {$\langle\,\mathbb{D}_{\text{c}}\rangle$};

  \draw (3,2.5) circle (0.8);
  \node at (2.8,2.5) {$\mathbb{D}$};

  \draw (4,2.5) circle (0.8);
  \node at (4.2,2.5) {$\mathbb{D}_{\text{c}}$};

\end{tikzpicture}
\caption{This Venn diagram classifies the literal form games that are equal to dyadic rationals; later we will refer to them all as numbers.}
    \label{fig:venntildeDzugzwang}
\end{figure}

In the previous Section~\ref{sec:dyadicalt}, while introducing the sets $\mathbb{D}_{\text{c}}, \langle \, \mathbb{D}_{\text{c}} \rangle$ and $\widetilde{\mathbb D}$, we discussed the importance of distinguishing between them.  Continuing that discussion, at the beginning of this section, we introduced another relevant set: the set ``$\zug$ '' of zugzwang games. We now have a complete picture of these distinctions, given by the Venn diagram in Figure~\ref{fig:venntildeDzugzwang}. These distinctions highlight another flaw in Siegel's book~\citep[Corollary 3.11]{S2013}. A corollary related to Proposition~\ref{prop:zugzwangequalD} states that: ``\textit{If a short game $x$ is equal to an S-number, then $x \in \mathbb{D}$}.''\footnote{Siegel defines the set $\mathbb{D}$ in the same way as we do~\citep[Definition 3.4]{S2013}, but he does not explicitly state the interpretation of the game form $m/2^k$.} 
However, consider the literal form game $\cg{*}{*}$.
\begin{enumerate}[i)]
\item Its simplest form is the S-number  $0$.
    \item But $\cg{*}{*}$ is not in $\mathbb D$.
\end{enumerate}
This contradicts the statement, showing that the corollary is incorrect. 

In this work, we distinguish between the statements ``$x$ is an S-number'' and ``$x$ is equal to an S-number'': the former means that $x$ satisfies the structural properties defining S-numbers, while the latter allows $x$ to be equivalent in game value without itself possessing those properties. Even if we would modify the corollary to state ``\textit{If a short game $x$ is an S-number, then $x \in \mathbb{D}$}'', it will still not hold. For instance, the game $\cg{1}{2}$ is an S-number but not in $\mathbb D$.  

Therefore, a more accurate version of the statement is: ``\textit{If a short game $x$ is an S-number, then $x$ is equal to dyadic, that is $x \in \widetilde{\mathbb{D}} $. }'', for which the original justification in the book remains valid. We establish this result via Lemma~\ref{lem:zugzwangfittingset} and Proposition~\ref{prop:zugzwangequalD}. Moreover, by the definition of $\widetilde{\mathbb D}$, the more general statement: ``\textit{If a short game $x$ is equal to an S-number, then $x \in \widetilde{\mathbb{D}}$}'', is also true. 

In this text, we have not yet defined what we mean by ``numbers''.   Among short games, only dyadic rationals behave as they do in the real number system.  Moreover, when we analyze games, we want to think about equvalence classes. Thus, we propose the following.

 \begin{defi}[Numbers]\label{def:numberasdalt}
A game $G$ is called a number if it belongs to $\widetilde{\mathbb D}$.
 \end{defi}


Given a game form, it can be tedious to compute its game value as it requires all its subpositions to be in canonical form.  In this spirit, when we inquire whether a game is a number (with relevance in for example Theorem~\ref{thm:numberavoidalt}), we have one more tool, related to the Simplicity Theorem, that can sometimes be more efficient.\footnote{We thank Prof. Carlos P. Santos for suggesting this alternative characterization.}

\begin{thm}
    A literal form game $G \in \mathbb G$ is  a number if and only if its fitting set $\mathcal F(G) \neq \emptyset$. 
\end{thm}

\begin{proof}
Suppose $G \in \mathbb{G}$ is a number. Then $G \in \widetilde{ \mathbb D}$, which means there exists  $x \in \mathbb D$ such that $G=x$. Since $x$ is also a zugzwang, it follows from the definition of $\mathcal{F}(G)$ that $x \in \mathcal{F}(G)$.  Therefore, $\mathcal{F}(G) \neq \emptyset$.

 For the converse, suppose for a game $G \in \mathbb G$, $\mathcal F(G) \neq \emptyset$. Then, by the Simplicity Theorem~\ref{thm:dyadsimplicityalt}, we have $G = x$, where $x$ is the simplest element of $\mathcal{F}(G) \subseteq \mathbb{D}$. Hence, $G$ is a number.
\end{proof}

We now establish an important property of the set $\mathbb{G}$, analogous to the Archimedean property of real numbers, which ensures that every game is bounded above and below by some integer.

\begin{thm}[Archimedean Property]\label{thm:archimedean} 
    For any game $G \in \mathbb{G}$, there exists an integer $n \in \mathbb{N}$ such that $-n<G<n$.
\end{thm}

\begin{proof}
    Consider a game $G \in \mathbb{G}$. Choose an integer $n > b(G)$. 
    
    Consider the game $n - G$. In this game, Right can make at most $n - 1$ moves, whereas Left has at least $n$ free moves in the game. Therefore, regardless of the starting player, Left can secure a win by playing all her moves in the $n$ component. Thus, $n - G > 0$, which implies $n > G$. 

    Similarly, consider the game $n + G$. Again, Right can make at most $n - 1$ moves, while  Left has at least $n$ free moves in the game. Therefore, Left can ensure a win regardless of who starts. Thus, $n + G > 0$, which implies $G > -n$.
\end{proof}

From the proof of Archimedean property Theorem~\ref{thm:archimedean}, it is easy to observe that for any $n \in \mathbb N$, if a game $G$ has birthday smaller than $n$, then the game $n>G$. 

 In canonical form, number games are zugzwangs, and in zugzwangs, no player has an incentive to move. Thus, if we are winning a game $G+x$, where $x$ is a number game, we expect that we do not start by playing on $x$. 
 
\begin{thm}[Number Avoidance Theorem]\label{thm:numberavoidalt} 
Suppose $x$ is a number game, and $G$ is a game that is not a number. If a player has a winning move in the sum $G+x$,  then they have a winning move in the $G$ component.
\end{thm}

\begin{proof}
    Suppose Left has a winning move in the game $G+x$. We will show that there exists a Left option $G^L$ such that $G^L + x \ge 0$. The argument for Right is analogous.

If Left already has a winning move in the $G$ component, there is nothing to prove. So assume her only winning move is in the $x$ component, that is, a move to $G + x^L$. This implies $G + x^L \ge 0$.
   
   Since  $x^L$ is a number and $G$ is not, it follows that $G \neq x^L$, and hence $G+x^L > 0$. Therefore, Left has a winning move in the game $G+x^L$. By induction, this winning move must be in the $G$ component, that is, there exists $G^L$ such that $G^L+x^L \ge 0$. Since $x$ is a number, we get $G^L+x > G^L+x^L\ge 0$. 

   Thus, Left has a winning move in the $G$ component of $G + x$.
\end{proof}








We conclude this section with an open problem and a related partial result. In our earlier definition of \emph{zugzwang}, we required that the inequality $G^L < G < G^R$ hold not only for the game $G$, but also recursively for all of its subpositions. This recursive condition made the definition strong. We now consider a relaxation of this condition by removing the requirement on subpositions. This gives rise to a weaker variant, which we term a \emph{weak zugzwang}. 

More precisely, a game $G = \cg{G^\mathcal{L}}{G^\mathcal{R}}$ is called a \emph{weak zugzwang} if for every $G^L \in G^\mathcal{L}$ and every $G^R \in G^\mathcal{R}$, we have $G^L < G < G^R$.

\begin{problem} \label{prob:nondyadiczugzwang}
Does there exist a non-dyadic weak zugzwang?
\end{problem}

To simplify our investigation, we restrict attention to canonical form games of the type $G = \cg{G^L}{G^R} \in \mathbb{G}$ that have only one option for each player.

We observe the following:

\begin{itemize}
    \item By Lemma~\ref{lem:dyadtildedensealt} and Simplicity Theorem~\ref{thm:dyadsimplicityalt}, if both $G^L$ and $G^R$ are dyadic, then $G$ is also dyadic.
    \item If $G^L < 0$ and $G^R > 0$, then $G = 0$.
\end{itemize}

Therefore, in order for $G$ to be a potential non-dyadic weak zugzwang, both $G^L$ and $G^R$ must lie on the same side of $0$, either both positive or both negative.

Further analysis shows that if $G^{LL} < G^L < G^R < G^{RR}$, then $G$ satisfies $G^L < G < G^R$, and hence is a weak zugzwang. However, we have not yet been able to construct an explicit example of a game with this property.

\section{Beside numbers}\label{sec:infinitesimalsalt}
To further investigate Problem~\ref{prob:nondyadiczugzwang}, we introduce another important class of combinatorial games known as \emph{infinitesimals}.  These are games whose values are smaller than every positive number and greater than every negative number, yet not necessarily equal to zero. More precisely:

\begin{defi}[Infinitesimals]
 A game $G \in \mathbb G$ is infinitesimal if $-x<G<x$ for every positive number $x \in \mathbb{D}$.
\end{defi}

To characterize infinitesimals, we now introduce the notion of \emph{dicotic} games:

\begin{defi}[Dicotic]
A game $G \in \mathbb G$ is called dicotic if both Left and Right have a move from every nonempty subposition of $G$.
\end{defi}

For example, the games $0$, $\cgstar = \cg{0}{0}$, and $\cgup = \cg{0}{\cg{0}{0}}$ are dicotic, but $1 = \cg{0}{\varnothing}$ is not. 

The following theorem provides a simple method for identifying infinitesimal games: given a game $G$, it suffices to check whether both players have options from every subposition.
\begin{thm} \label{thm:dicoticinfinitesimal}
Every dicotic game is an infinitesimal.
\end{thm}

\begin{proof}
    Suppose $G$ is a dicotic game and let $x$ be an arbitrary positive number. To prove that $G$ is an infinitesimal, we need to show that $x>G>-x$. It suffices to prove that Left wins the game $x-G$ irrespective of who starts. The proof that Right wins  $-x-G$ irrespective of who starts is analogous. 

    If $G$ is a zero game, then there is nothing to prove. So, assume that $G$ is non-zero.

    Consider the game $x-G$. If Left starts, she plays to $x-G^R$. Since $G$ is dicotic and non-zero, it must have at least one Right option. By induction, Left wins the game $x - G^R$.
    
    If Right starts, he plays to either $x-G^L$ or to $x^R-G$. In the first case, by induction, Left wins $x-G^L$. In the second case, since, $x^R>x>0$, Left responds by playing in the $G$ component to $x^R-G^R$. By induction, she wins.

    Therefore, Left always wins $x - G$, and it follows that $G < x$. A similar argument shows that Right always wins the game $-x - G$, which implies $G > -x$. Hence, $G$ is an infinitesimal. 
\end{proof}

By Theorem~\ref{thm:dicoticinfinitesimal}, it is immediate that the games $\cgstar$ and $\cgup$ are infinitesimals, as they are dicotic.

Since every dicotic game is infinitely small compared to numbers, they are also called \textbf{all-small} games.

However, not all infinitesimals are dicotic. There exist infinitesimal games that are not all-small.

\begin{ex}
Consider the game $\cgtiny_1 = \cg{0}{\cg{0}{-1}}$. This game is not dicotic, as it has a Right option to $-1$, from which only Right can move, violating the dicotic condition. Since $0 < \cgtiny_1 < \cgup$, the game $\cgtiny_1$ is still an infinitesimal.
\end{ex}

We now return to our investigation of Problem~\ref{prob:nondyadiczugzwang} from the previous Section~\ref{sec:numbersalt}, of finding a non-dyadic game $G = \cg{G^L}{G^R}$ such that $G^L<G<G^R$. Earlier, we observed that if a game $G = \cg{G^L}{G^R}$ satisfies $G^{LL}<G^L<G^R<G^{RR}$, then it follows that $G^L<G<G^R$. With this aim, we now explore the relation of some positive infinitesimals with their Left and Right options. See Table~\ref{tbl:infinitesimals}.

\begin{table}[ht]
\centering
\renewcommand{\arraystretch}{1.5}
\caption{Comparison of positive infinitesimals with their Left and Right options.}
\begin{tabular}{|c|c|c|}
\hline
 $G = \cg{G^L}{G^R}$ & Relation with $G^L$ & Relation with $G^R$ \\ \hline
$\cgstar = \cg{0}{0}$ & $0 \cgfuzzy \cgstar$ & $0\cgfuzzy\cgstar$\\ \hline
$\cgup = \cg{0}{\cgstar} $ & $0<\cgup $ & $ \cgstar \cgfuzzy \cgup$ \\ \hline
$\cgup\cgstar = \cg{0, \cgstar}{0}$ & $0\cgfuzzy \cgup\cgstar$, $\cgstar < \cgup \cgstar$ & $0 \cgfuzzy \cgup\cgstar$ \\ \hline
$\cgdoubleup = \cg{0}{\cgup\cgstar}$ & $0 < \cgdoubleup$ & $\cgup\cgstar \cgfuzzy \cgdoubleup$ \\ \hline
$\cgdoubleup \cgstar = \cg{0}{\cgup}$ & $0<\cgdoubleup \cgstar$ & $\cgup \cgfuzzy \cgdoubleup \cgstar$ \\ \hline
$\cgtiny_1 = \cg{0}{\cg{0}{-1}}$ & $0< \cgtiny_1$ & $\cg{0}{-1} \cgfuzzy \cgtiny_1$ \\ \hline
\end{tabular}
\label{tbl:infinitesimals}
\end{table}

From Table~\ref{tbl:infinitesimals}, we observe a consistent pattern: in each of the positive infinitesimals considered, the Right option is always fuzzy with the game. This raises the following question:

\begin{problem} Does there exist a positive infinitesimal $G = \cg{G^L}{G^R}$ such that $G < G^R$? \end{problem}

If such a positive infinitesimal exists, that is, one satisfying $G < G^R$, then, combined with the fact that some positive infinitesimals satisfy $G^L < G$, we could construct a game $G = \cg{G^L}{G^R}$ where both options are infinitesimals and satisfy $G^{LL} < G^L < G^R < G^{RR}$. By Lemma~\ref{lem:GLGGR}, this would then give us a non-dyadic positive infinitesimal game with the weak zugzwang property,  $G^L < G < G^R$. 

Since no such infinitesimal with $G < G^R$ has been found among those we examined, we turn to an alternative approach. We consider games where the Right option is dyadic, as dyadics have the property that their Right option is strictly greater than the game itself. Therefore, we explore a hybrid configuration: games $G = \cg{G^L}{G^R}$ where $G^L$ is a positive infinitesimal satisfying $G^{LL} < G^L$, and $G^R$ is a dyadic satisfying $G^R < G^{RR}$. Such configurations result in games with the weak zugzwang property. Examples of such configurations are presented in Table~\ref{tbl:weakzugzwang}.

\begin{table}[ht]
\centering
\renewcommand{\arraystretch}{1.5}
\caption{Examples of literal form weak zugzwangs with dyadic game values.}
\begin{tabular}{|c|c|c|c|}
\hline
$G = \cg{G^L}{G^R}$ & Relation with $G^L$ & Relation with $G^R$ & game value \\ \hline
$G_1 = \cg{\cgup}{1/2}$ & $\cgup < G_1$ & $1/2>G_1$ & $G_1 =1/4$ \\ \hline
$G_2 = \cg{\cgdoubleup}{1/4}$ & $\cgdoubleup<G_2$ & $1/4>G_2$ & $G_2 =1/8$ \\ \hline
$G_3 = \cg{\cgdoubleup\cgstar}{1/8}$ & $\cgdoubleup \cgstar<G_3$ & $ 1/8>G_3$ & $G_3 = 1/16$ \\ \hline
$G_4 = \cg{\cgtiny_1}{1/16}$ & $\cgtiny_1<G_4$ & $1/16>G_4$ & $ G_4 = 1/32 $\\ \hline
\end{tabular}
\label{tbl:weakzugzwang}
\end{table}

As shown in Table~\ref{tbl:weakzugzwang}, each of these games satisfies the weak zugzwang property, but their game values are still dyadic. This occurs because the dyadic rationals are dense, ensuring that the fitting set $\mathcal{F}(G_i)$ is nonempty for each game $G_i$. Consequently, we cannot construct a non-dyadic weak zugzwang where either $G^L$ or $G^R$ is dyadic. This leads to the following conjecture:

\begin{conj} 
If a game $G$ is a weak zugzwang, then $G$ is equal to a dyadic.
\end{conj}

\subsection*{Epilogue}

Our recent work on Bidding Combinatorial Games~\citep{KLRU,kant2025constructive, kant2025thesis} generalizes alternating normal play games. While developing this area, we discovered that number games exhibit nice properties.  
To build further on this for our bidding setting, we revisited the foundations of numbers in alternating play~\citep{ONAG, S2013, WW}. To our surprise, we encountered certain gaps that appear to have gone unnoticed in the literature; in particular, the assurance of non-emptiness of the interval $\mathcal I$ in the Simplicity Theorem for numbers~\citep[Corollary 3.11]{S2013} had not been properly addressed. Among these gaps, a major issue was the lack of distinction between a game being a number and a game being equal to a number.

In this work, we  fill those gaps by redefining numbers and establishing the notion of a ``fitting set''. However, through this process, we also realized that the fitting set approach, while powerful in alternating play, does not naturally extend to general bidding games. A follow-up paper, currently in preparation, will explore how the theory of numbers may  evolve in the bidding setting.

\bibliographystyle{plain}
\bibliography{ref}

\end{document}